\documentclass[aip,
jcp,%
amsmath,amssymb,
reprint,%
]{revtex4-1}

\usepackage{hyperref}
\pdfoutput=1
\usepackage{graphicx}
\usepackage{amssymb}
\usepackage{amsmath}
\usepackage{blkarray}
\usepackage{multirow}
\usepackage{natbib}
\usepackage{epstopdf}
\usepackage{float}

\usepackage{graphicx}
\usepackage{dcolumn}
\usepackage{bm}

\usepackage{tabu} 

\begin{document}

\title{Dynamics and thermodynamics of Ibuprofen conformational isomerism at the crystal/solution interface.}

\author{Veselina Marinova}
\affiliation{Thomas Young Centre and Department of Chemical Engineering, University College London, London WC1E 7JE, UK.}%
\author{Geoffrey P. F. Wood}
\affiliation{Pfizer Worldwide Research and Development, Groton Laboratories, Groton, Connecticut 06340, USA}%
\author{Ivan Marziano}
\affiliation{Pfizer Worldwide Research and Development, Sandwich, Kent CT13 9NJ, UK}%
\author{Matteo Salvalaglio}%
\email{m.salvalaglio@ucl.ac.uk}
\affiliation{Thomas Young Centre and Department of Chemical Engineering, University College London, London WC1E 7JE, UK.}%

\date{\today}

\begin{abstract}
Conformational flexibility of molecules involved in crystal growth and dissolution is rarely investigated in detail, and usually considered to be negligible in the formulation of mesoscopic models of crystal growth. 
In this work we set out to investigate the conformational isomerism of ibuprofen as it approaches and is incorporated in the morphologically dominant \{100\} crystal face, in a range of different solvents - water, 1-butanol, toluene, cyclohexanone, cyclohexane, acetonitrile and trichloromethane. 
To this end we combine extensive molecular dynamics and well-tempered metadynamics simulations to estimate the equilibrium distribution of conformers, compute conformer-conformer transition rates,  and extract the characteristic relaxation time of the conformer population in solution, adsorbed at the solid/liquid interface, incorporated in the crystal in contact with the mother solution, and in the crystal bulk.  
We find that, while the conformational equilibrium distribution is weakly dependent on the solvent, relaxation times are instead significantly affected by it. Furthermore, differences in the relaxation dynamics are enhanced on the crystal surface, where conformational transitions become slower and specific patways are hindered. This leads to observe that the dominant mechanisms of conformational transition can also change significantly moving from the bulk solution to the crystal interface, even for a small molecule with limited conformational flexibility such as ibuprofen.  Our findings suggests that understanding conformational flexibility is key to provide an accurate description of the solid/liquid interface during crystal dissolution and growth,  and therefore its relevance should be systematically assessed in the formulation of mesoscopic growth models.

\end{abstract}

\keywords{conformers, crystal growth, Markov State Models, metadynamics}
\maketitle

\section{\label{sec:intro}Introduction}
Cystallisation is a key step in the production of active pharmaceutical ingredients (APIs) as it can determine their purity, bulk structure, and size and shape distributions. Crystal habits of APIs are known to affect their mechanical properties and bulk density \cite{HadiA.GarekaniJamesL.FordMichaelH.Rubinstein1999,Garekani2001}, as well as their dissolution rates and consequently, their bioperformance\cite{Hooton2008,N.BlagdenM.deMatasP.T.Gavan2007}. The shape of crystal particles also impacts their compaction properties \cite{P.V.Marshall1991} and handling in powder form with flakes and needles being the most difficult to process \cite{SoojinKim2003}. A number of factors affect crystal growth from solution such as, the identity of the solvent, presence of impurities and cooling rate. The modification of the crystal habit of organic molecules by changing the solvent has been extensively studied \cite{Sudha2014,DaveyR.J.MullinJ.W.Whiting1982,Bunyan1991}, stressing the importance of crystal habit prediction for optimal process and product design \cite{Variankaval2008}. 

The shape of a crystal grown from solution is determined by the relative growth rate of individual crystal faces exposed to the solvent. Crystal faces of great morphological importance grow slowly compared to other faces, whilst faces characterised by a small surface area are fast-growing. Being able to predict crystal shapes depends on obtaining reliable information on the growth kinetics of individual faces, either through molecular simulations or energy approximations in mathematical models \cite{Lovette2008}. Over the last few decades, significant progress has been made in the accurate prediction of solution grown crystal shapes and a number methods have been developed, some of which have the capability of accounting for solvent effects. The current state of the art in this field comprises of approaches relying on mechanistic models or molecular simulations. In particular, the most recent contribution in the area of mechanistic models is represented by the work of J. Li and M. Doherty \cite{Li2017}. In their work, the authors introduce a mechanistic spiral growth (MSG) model to predict the steady state morphologies of a paracetamol crystal, grown from 30 different solvents. The method comprises of kink rate and kink density calculations, as well as solvent interaction calculations. The predicted morphologies match the experimental shapes very well for the cases of vapour and water growth, demonstrating the accuracy of the theory. In turn, molecular dynamics (MD) simulations, are invaluable for exploring the dynamic evolution of systems of interest, though at times the domain of applicability of MD can be limited by sampling. 

\begin{figure}[h!]
	\includegraphics[width=1.0\linewidth]{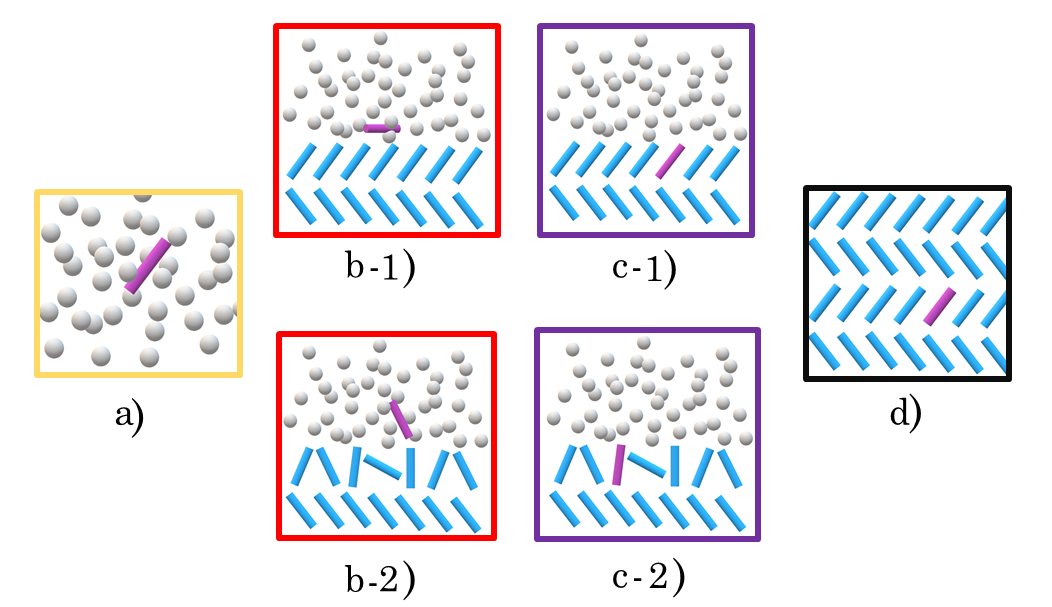}
	\centering
	\caption{ Scheme of states of an ibuprofen molecule along the process of incorporation in the crystal bulk - in the bulk solution (yellow frame), surface adsorbed, on an ordered or disordered surface (red frame), within the outer crystal surface layer of an ordered or disordered surface (purple frame), in the bulk of the solid (black frame).}
	\label{fig:states_definitions}

\end{figure}

In current crystal shape prediction methods, conformational flexibility is typically neglected. In fact, a general approach to the systematic incorporation of conformational flexibility in crystal growth models is still missing\cite{Elts2017}. 
In this work we aim to highlight how, even for relatively small molecules, conformational flexibility and crystal growth/dissolution are inherently coupled. To tackle this problem we analyse the conformational flexibility of a single ibuprofen molecule in four states along its path of incorporation into the crystal bulk. We choose to focus on a single molecule as a conformationally flexible unit rather than making an assumption on the nature of the growth unit involved in the process.
The four states are depicted in ~\figurename{~\ref{fig:states_definitions}}) and include: bulk solution, surface adsorbed, within the outer crystal surface layer and in the bulk of the solid. 
For all states we investigate how the solvent, as well as the structure of a crystal/solution interface impacts ibuprofen conformational transition kinetics, thermodynamics and mechanisms. Our work paves the way towards a formulation of solvent-dependent kinetic models for the growth and dissolution of flexible molecules, necessary to predict crystal shape evolution in systems of practical relevance.

\section{\label{sec:methods}Methods}
In the present study we investigate the conformational flexibility of ibuprofen in four different states as described in ~\figurename{~\ref{fig:states_definitions}} using all-atom MD simulations. Since results from MD are only meaningful when all energetically relevant configurations have been explored, in certain cases, MD is combined with well-tempered metadynamics (WTmetaD) \cite{Barducci2008} for computational efficiency. State-to-state kinetics and conformational rearrangement mechanisms are then reconstructed within the theoretical framework of Markov State Models (MSM) \cite{Bowman2014}. 

\subsection{Simulation setup }

\begin{figure}[h!]
	\includegraphics[width=1.0\linewidth]{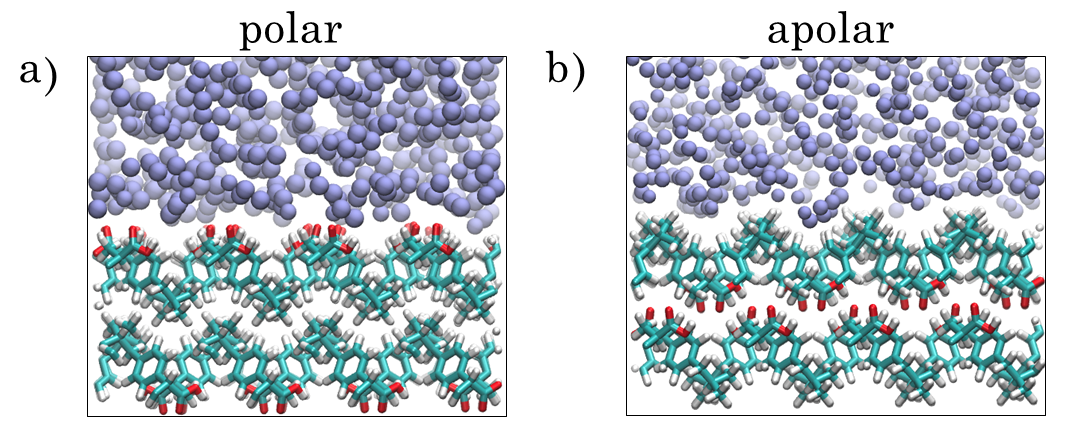}
	\centering
	\caption{\textbf{a)} Ibuprofen \{100\} polar surface (left) and \textbf{b)} ibuprofen \{100\} apolar surface (right), where solvent molecules are represented as spheres for simplicity. }
	\label{fig:surface_layers}

\end{figure}

MD simulations were performed with Gromacs 5.1.4  \cite{VanDerSpoel2005} using the Generalised Amber Force Field (GAFF) \cite{Wang2004} and an explicit representation of the solvent. Force field parameters for solvent molecules were obtained from the Virtual Chemistry solvent database \cite{Caleman2012,VanderSpoel2012}. Where relevant, WTmetaD and collective variable post-processing have been carried out using Plumed 2.3 \cite{Tribello2013} . \\
\indent
Seven solvents were considered in this study - water, 1-butanol, toluene, cyclohexanone, cyclohexane, acetonitrile and trichloromethane. The conformational complexity of the ibuprofen molecule was considered for each solvent in a surface state, an adsorbed state and in solution. In our study we have considered the morphologically dominant \{100\} crystal face. 
Due to the layered character of the \{100\} face and depending on where a slice is made, either polar or apolar functionalities may be exposed to the external environment, as shown in ~\figurename{~\ref{fig:surface_layers}}.  In the present study, we investigate both scenarios for each solvent, as discussed in detail in the results section.

The kinetic and thermodynamic information in each case was recovered using independent simulations. In addition, kinetics recovered with the aid of metadynamics involved performing 50 simulations per starting configuration.  For a detailed summary of all simulations performed please refer to Section A of the Supplementary Material. 

\paragraph*{Molecular Dynamics Setup}
Prior to MD each system undergoes an energy minimisation using a conjugate gradient algorithm with a tolerance on the maximum residual force of 100 kJ mol$^{-1}$ nm$^{-1}$. 
In all MD simulations a cut-off of 1.0 nm for non-bonded interactions is chosen, three-dimensional periodic boundary conditions (\textit{pbc}) are applied and long range intermolecular interactions are included using the particle-mesh Ewald (PME) approach \cite{Darden1993}. A time step of 2 fs was used in conjunction with constraining the fast bond vibrations by applying the LINCS algorithm \cite{Hess1997}. All of the simulations analysed in this work were carried out within the isothermal-isobaric (NPT) ensemble at pressure of 1 bar and temperature of 300 K, using Bussi-Donadio-Parrinello thermostat \cite{Bussi2007} and Berendsen barostat \cite{Berendsen1984}. Coupling between the x, y and z components in the barostat setup is varied based on the environment in which the conformational flexibility of ibuprofen is investigated. A fully isotropic pressure control is used for bulk solution simulations. For simulations in the crystal bulk, we apply an anisotropic pressure scaling, in order to explicitly account for any potential deformation of the crystal lattice induced by local conformational transitions. Simulations of the solid/liquid interface are carried out by using a semi-isotropic pressure control, as it allows the continuous adaptation of the system volume by scaling only the z-coordinate, avoiding rescaling distances on the x,y plane, which could destabilise the crystal slab. 

\begin{figure}[h!]
	\includegraphics[width=1.0\linewidth]{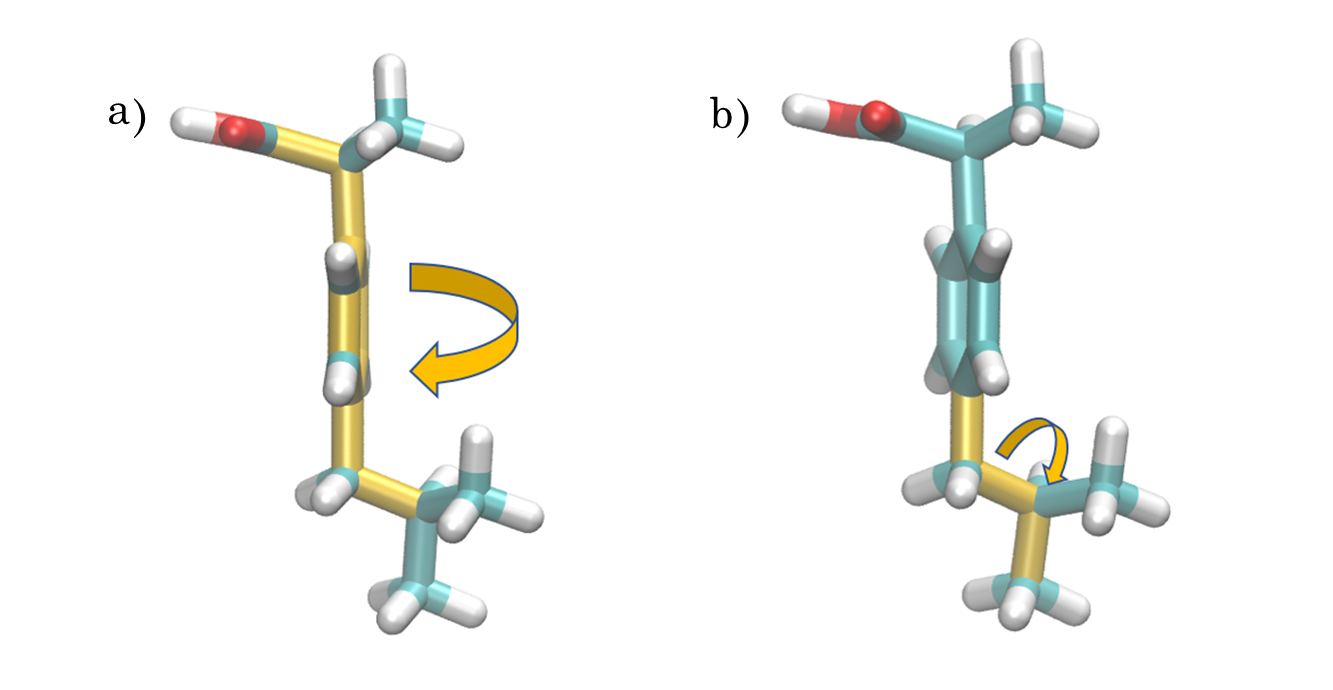}
	\centering
	\caption{Internal torsional angles of ibuprofen used to map the conformational space of the molecule. a) Global torsional angle monitors the rotation of the methyl groups around the vertical axis of the molecule and b) Local torsional angle monitors the degree of interchange between the methyl groups.}
	\label{internal_torsions}
\end{figure}

\paragraph*{Collective Variables} 
The conformational ensemble of ibuprofen is mapped on a two-dimensional collective variable (CV) space, defined by two internal torsional angles that represent a \emph{global} and \emph{local} rearrangement of the ibuprofen molecular structure.  
As shown in Fig. \ref{internal_torsions}, the global torsion captures the relative orientations of the \emph{para}-substituents with respect to one another around the plane of the phenyl ring, whereas the local torsion captures the relative rotational angle of the methyl groups of the isobutanyl substituent. 

For a quantitative assessment of the crystal-solution interface structural rearrangement  we define an order parameter (SMAC) that captures the local order within the coordination sphere of an ibuprofen molecule. SMAC is based on the evaluation of the density of the first coordination shell and on the relative orientation of neighbours, considering the bulk crystal structure as a reference. Details for the mathematical definition of SMAC can be found in Ref. \cite{Giberti2015} and \cite{Tribello2017}. The SMAC parameters used for ibuprofen in the present study are reported in Section B in the Supplementary Material. 

\begin{figure*}[ht]
	\includegraphics[width=0.85\linewidth]{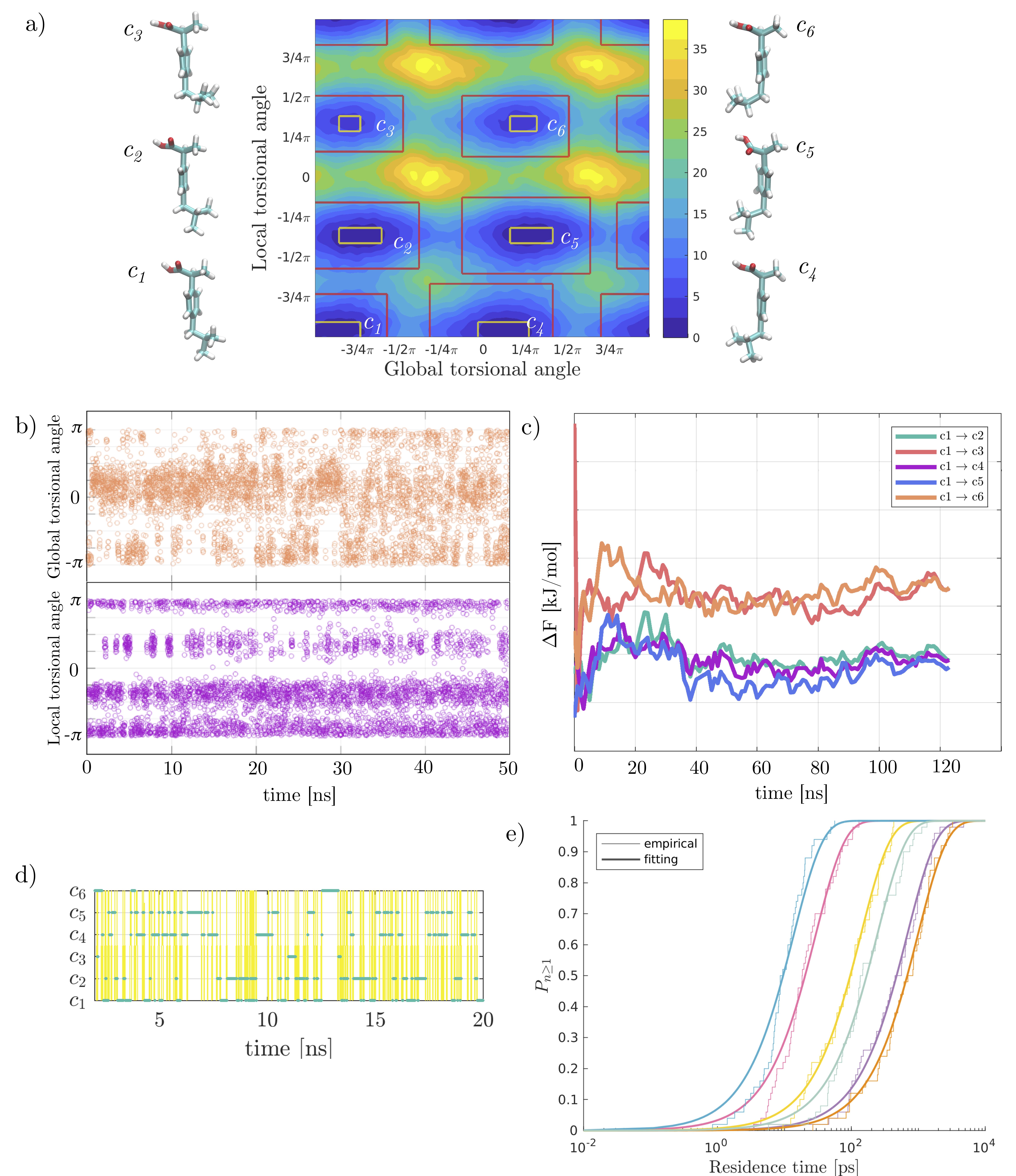}
	\centering
	\caption{\small{\textbf{a)} Free energy surface in the conformational space of the two internal torsional angles of ibuprofen showing the integration domain used for calculating equilibrium distributions from WTmetaD in red and the core set domain in yellow. Molecular structures of the conformers identified by the local minima are reported alongside the FES. \textbf{b)}  Global (orange) and Local (purple) torsional angle vs. time in a WTmetaD simulation \textbf{c)} Convergence of free energy differences of conformers with respect to c1. \textbf{d)} State to state transitions in an unbiased MD simulation. \textbf{e)} Conformer lifetime distribution for an ibuprofen molecule in the apolar \{100\} surface in water, recovered with the aid of WTmetaD.}}
	\label{panel1}
\end{figure*}

\paragraph*{WTmetaD Setup}
WTmetaD has been used as a tool to increase the sampling efficiency of ibuprofen conformational space \cite{Barducci2008}, as shown for a representative simulation trajectory in Fig. ~\ref{panel1}b). This allows us to compute free energy surfaces (FES) in CV space, and to estimate the rate of rare conformational transitions \cite{Tiwary2013}.
In WTmetaD simulations an external time-dependent biasing potential is introduced as a function of the torsional angle CVs discussed earlier.  The external bias potential is defined as a sum of Gaussian functions. The height and width of the Gaussian functions, the pace of bias update and the bias factor are the parameters controlling the construction of the bias potential \cite{Barducci2011}. In Tab. \ref{metad_table} we report the parameters used for simulations to construct the free energy surface of the conformational space of ibuprofen, as well as for simulations used to calculate rates of conformational transitions.

\begin{table}[!h]
\caption{Parameters in WTmetaD simulations, used to compute free energy surfaces (FES) and conformational transition rates.}
	\begin{tabular}{ c | cccc }
		& \textbf{Width} & \textbf{Height} & \textbf{Bias Factor} & \textbf{Pace} \\
		& [ rad ] & [$k_BT$] & [K] & [steps] \\
		\hline

		FES &  0.1 & 0.96 & 10 & 500  \\		
		Rates &0.1 & 0.24 & 5 & 1500 \\ 
	\end{tabular}
		\label{metad_table}
\end{table}

\paragraph*{Conformational equilibrium probability distribution from WTmetaD}

The FES computed in collective variables space $F(\mathbf{S})$ provides information on the equilibrium probability density $p(\mathbf{S})$, defined as: 
\begin{equation}
p(\mathbf{S})=\frac{e^{-\beta{F(\mathbf{S})}}}{{\int_\Omega e^{-\beta{F(\mathbf{S})}} d\mathbf{S}}}
\label{probability_density}
\end{equation}
where $\Omega$ indicates the entire configuration space, and $\beta^{-1}=k_BT$, where $k_B$ is the Boltzmann constant. 
The FES projected in the space of the global and local torsional CVs can therefore be used to compute the equilibrium probability of different conformers. 
For instance, as shown in the Fig. ~\ref{panel1}a), a typical FES displays a set of local minima. Each minimum corresponds to the projection in CV space of an ensemble of configurations that can be identified as a single ibuprofen conformer. Hence, by integrating the probability density $p(\mathbf{S})$ over the sub-domain corresponding to a local minimum in CV space, one can obtain the equilibrium probability of the conformer corresponding to that minimum. For the generic conformer $i$, the equilibrium probability $P_i$ is computed as: 
\begin{equation}
P_i=\int_{\Omega_i} p(\mathbf{S}) d\mathbf{S}n
\label{probability_state}
\end{equation}
where $\Omega_i$ indicates the region of CV space pertaining to conformer $i$. The integration domains of each conformer have been indicated in Fig. \ref{panel1}a) in red. 
The free energy difference between any pair of $i$ and $j$ conformers is therefore computed as $\Delta{F}_{i,j}=-\beta^{-1}\log{\frac{P_j}{P_i}}$. The convergence of WTmetaD simulations has been systematically assessed by monitoring free energy differences between conformers as a function of simulation time, as shown in Fig. \ref{panel1}c). Each simulation is considered converged once the fluctuations in the free energy between the minimum corresponding to conformer c1 and all other conformers are less than 5 kJ mol$^{-1}$. In the cases in which the sampling of the conformational space has been restricted through the application of a static bias potential, an appropriate reweighting of the results has been carried out. 

\paragraph*{Kinetics from WTmetaD}
Performing metadynamics with the aim of obtaining conformational transition kinetics requires constructing bias to favour the escape from a local free energy minimum without perturbing the transition state ensemble. This is typically achieved by reducing the pace of bias construction\cite{Tiwary2013}, and by comparing the shape of the transition times distribution with that of an exponential distribution implied by the \emph{law of rare events}\cite{Salvalaglio2014}, as shown in ~\figurename{~\ref{panel1}}e). In this work we carry out this process by constructing distributions of transition times from 50 events obtained from independent WTmetaD simulations. 
Results of the Kolmogorov-Smirnov test, employed to analyse the shape of the distribution of transition times, are reported in Tab. IV in the Supplementary Material. 


\subsection{Construction of a Markov State Model}
\paragraph{Equilibrium probability and system relaxation time}
In this work, we are interested in using a Markov State Model (MSM) to compute the global relaxation rate of the conformational population, the mechanism of conversion between conformers, and their equilibrium populations. The equilibrium populations obtained from metadynamics have been used to cross-validate the MSM (see the Supplementary Material, Section F).
An MSM is defined by a master equation of the following form:
\begin{equation}
\dot{\mathbf{P}}(t)=\mathbf{KP}(t)
\label{master_equation}
\end{equation}
where $\textbf{P}(t)$ is the vector with the probabilities of each macrostate at time $t$, and \textbf{K} is a transition rate matrix which has off-diagonal elements $k_{ij}\geq0$ and diagonal elements $k_{ii}=-\sum_{j\neq i}k{ji}<0$\cite{noe2014introduction,Bowman2014}.

Elements of the transition rate matrix can be computed either from unbiased simulations or from metadynamics, depending on the inherent timescale of the process\cite{Palazzesi2016}. 
In the case of unbiased simulations we apply a lifetime-based estimate \cite{Buchete2008} of the transition rate from state $i$ to state $j$:  $k_{ij}={\tau_i}^{-1}{n_{ij}}$, in which $\tau_i$ is the total residence time in state $i$ and $n_{ij}$ is the total number of transitions from state $i$ to state $j$. 
Transitions were accounted for only when the so-called core set of state $j$ is reached.  By doing so, we avoid including non-Markovian transitions in our analysis\cite{Buchete2008}. Core sets for all the ibuprofen conformational isomers are shown in \figurename{~\ref{panel1}}a) in yellow. A representation of state-to-state transitions within a typical trajectory is shown in \figurename{~\ref{panel1}}d).
For the cases where it is more practical to obtain kinetic information from metadynamics for computational efficiency, individual simulations starting from each conformer where carried out. The distribution of transition times was then compared to an exponential distribution, assessing their compatibility with a Kolmogorov-Smirnov test at 95 \% significance \figurename{~\ref{panel1}}e). 

Eq.~\ref{master_equation}, admits a unique stationary equilibrium distribution, determined by\cite{Buchete2008,noe2014introduction,Bowman2014}:
\begin{equation}
\textbf{KP}_{eq}=0
\label{eqilibrium_dist}
\end{equation}
\noindent
Solving for the eigenvalues and eigenvectors of the transition rate matrix \textbf{K} allows us to extract equilibrium probabilities of the conformational isomers $\mathbf{P}_{eq}$ and relaxation times. The eigenvalues of the matrix \textbf{K} can be sorted by magnitude: $\lambda_1=0 > \lambda_2 \geq \lambda_3 \geq ... \geq \lambda_N$. The equilibrium population is obtained from the normalization of the left eigenvector corresponding to $\lambda_1=0$. The largest non-zero eigenvalue of the \textbf{K} matrix corresponds to the relaxation time of the slowest mode of the system.

\paragraph{Transition pathway between macrostates} 
An MSM is a powerful tool for studying the most likely transition pathway between any two macrostates of the system. To this end, a knowledge of the transition probability matrix $\mathbf{T}$ between all miscrostates is required. Elements of the transition probability matrix $T_{ij}$ can be estimated from MD simulation by computing the number of transitions between each pair of states $i$ and $j$,  $C_{\textit{ij}}$:  
\begin{equation}
T_{ij}=\frac{C_{ij}}{\sum_{k=1}^N{C_{ik}}}
\label{transition_prob}
\end{equation}
where  $k$ represents each of the N macrostates.
\newline

With the knowledge of the transition probability matrix, a discrete transition pathway between any two states $A$ and $B$ is obtained by computing the committor probability \textit{$q{_n}^{+}$}. The latter is the probability that, when in the generic state \textit{n}, which belongs to the ensemble of all intermediate states \textit{I},  the system will visit state \textit{B} before state \textit{A}.
The committor probability \textit{$q{_n}^{+}$} is computed as \cite{Bowman2014}: 
 \begin{equation}
 -\textit{$q_{n}^{+}$} + \sum_{k \epsilon I} T_{nk}\textit{$q_{k}^{+}$}= - \sum T_{nB} \text{    for   } n \epsilon I
 \label{committor_calc}
 \end{equation}
where \textit{k} runs over the ensemble of all intermediate macrostates of the system.
By definition, the committor probability in the starting state A is 0 and in the final state B is 1. 

Once the committor probability for all states along the the discrete transition pathway between \textit{A} and \textit{B} is obtained, the net state-to-state probability flux between any two states $i$ and $j$, $f_{ij}^{+}$, is calculated using Eq.~\ref{flux} \cite{Bowman2014}:
\begin{equation}
\textit{$f_{ij}^{+}$}= \pi_{i} T_{ij}(\textit{$q_{j}^{+}$}-\textit{$q_{i}^{+}$})
\label{flux}
\end{equation}
The net flux and the committor probability allow us to quantitatively identify transition pathways and to infer model-based mechanistic hypotheses, reported in the Discussion section of the paper.  
\section{\label{sec:results}Results}
\begin{figure*}[!]
	\makebox[\textwidth][c]{\includegraphics[width=0.65\linewidth]{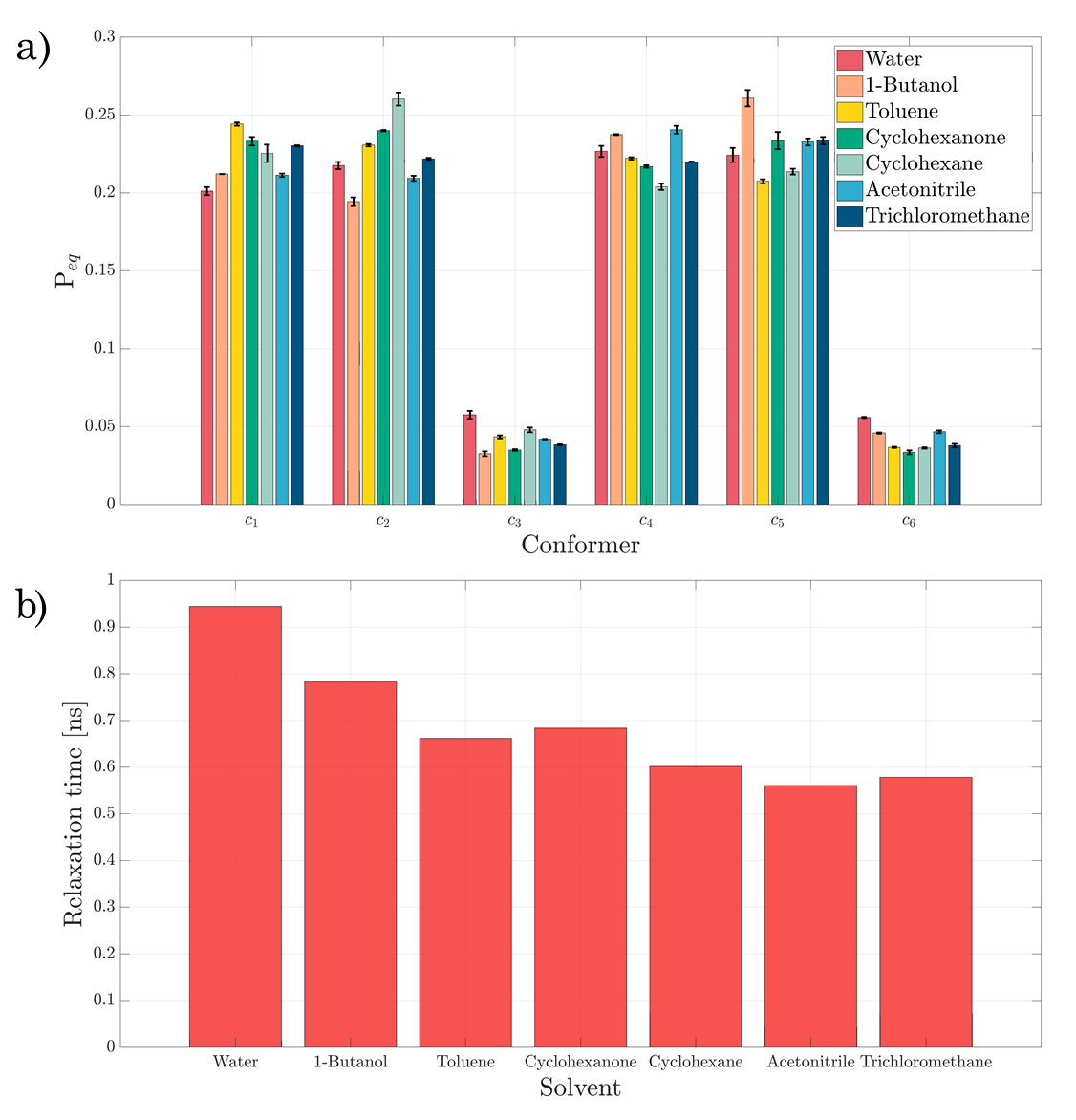}}
	\centering
	\caption{a) Equilibrium distribution of ibuprofen conformers in solution of different solvents recovered from WTmetaD. The solvents are shown in the order: water, 1-butanol, toluene, cyclohexanone, cyclohexane, acetonitrile, trichloromethane. Equilibrium distributions of ibuprofen conformers recovered from DFT/COSMO calculations and from MSMs are reported in the Supplementary Material (Fig. 3, Fig. 6, Fig. 7 and Fig. 8). b) System equilibration times for an ibuprofen in solution obtained from MSM analysis.  }
	\label{fig:solution_metad}

\end{figure*}

In this section we report thermodynamic and kinetic results as well as the mechanism for the conformational transitions of ibuprofen in solution and at the morphologically dominant \{100\} face.  In determining the crystal face with largest morphological importance we rely on BFDH model calculations, obtained using the software Mercury \cite{Macrae2008}, as well as experimental studies of solvent effects on the shape of ibuprofen crystals \cite{Bunyan1991,Shariare2012}. An extensive number of simulations (Supplementary Material, Section A) were performed in seven different solvents varying in size, polarity, and hydrogen bonding capabilities. The solvents considered are water, 1-butanol, toluene, cyclohexanone, cyclohexane, acetonitrile and trichloromethane. 
\newline 
\indent
To assess the effect of the presence of a crystal/solution interface we investigate ibuprofen conformational transitions in the four states mentioned in the introduction and sketched in ~\figurename{~\ref{fig:states_definitions}}. 
We start by introducing the results obtained for the equilibrium distribution and relaxation kinetics of ibuprofen conformers in bulk solution, as shown in ~\figurename{~\ref{fig:states_definitions}}a). In this case a single ibuprofen molecule is surrounded by solvent molecules in the absence of a crystal surface. Next, we discuss the equilibrium population of ibuprofen conformers found in the crystal bulk as shown in ~\figurename{~\ref{fig:states_definitions}}d). We complete the results on the dynamics and thermodynamics of conformational transitions with reporting the states at the crystal/solution interface, such as those sketched in Fig. ~\ref{fig:states_definitions}b, c). We conclude with a discussion on the mechanisms of conformational transitions obtained from the MSMs.

\subsection{Ibuprofen in bulk solution}
\subsubsection*{Equilibrium distribution of conformers in solution}
To  assess the impact of  solvent on the equilibrium probability of ibuprofen conformers in solution we carried out WTmetaD simulations of an isolated ibuprofen molecule, as described in the methods section. 
Such simulations uncovered six stable conformers present in each of the seven solvents investigated. The free energy surface for the case of an ibuprofen molecule in water is shown in \figurename{~\ref{panel1}}a) with the conformers labelled c1 through to c6. A comprehensive summary of all the free energy surfaces computed, accompanied by an analysis of their convergence, is reported in Fig. 9 and Fig. 10 of the Supplementary Material. 

The equilibrium probability of conformers in solution, computed using Eq. \ref{probability_density} and \ref{probability_state},  shows that conformers c1, c2, c4 and c5 dominate the distribution.  Each of these conformational isomers contribute to approximately one fifth of the overall distribution. The remaining two conformers, c3 and c6, are significantly less probable, each accounting for 5\% of the total conformer population. The lower probability of the latter two conformers can be accounted for by considering the position of the methyl groups of the iso-butanyl substituent with respect to the phenyl ring. Whilst in conformers c1, c2, c4 and c5 one of the methyl substituents is always pointing away from the molecule, in line with the central axis which contains the phenyl ring, in the case of conformers c3 and c6, the two groups are both equally close to the phenyl ring. This observation suggests that conformers c3 and c6 are much less probable due to the created steric effect. 
Small, but significant variations between the equilibrium probabilities of each conformer outside the calculated error are observed, indicating a weak dependence of the conformational distribution on the solvent. 
The equilibrium distribution of conformers identified from metadynamics was independently confirmed though a genetic search for stable structures \cite{Lauer2016}, coupled with free energy calculations, performed with internal gas-phase free energies calculated using Gaussian 09\cite{frisch2009gaussian} in conjunction with the PBE0/6-31G(d) functional, along with solvation free energies derived from COSMO\cite{Klamt2018} calculations. The number of stable ibuprofen conformers and their equilibrium probability found with the DFT/COSMO procedure are in qualitative agreement with the results obtained with WTmetaD, indicating that the collective variables chosen allow to exhaustively explore the internal degrees of freedom of the molecule and that the forcefield chosen for this study does not introduce significant artefacts. 

\subsubsection*{Relaxation Dynamics of the conformer population}
Unbiased MD simulations of an isolated ibuprofen molecule in different solvents allowed us to compute state-to-state transition probabilities and  construct a Markov State Model. Stable conformational states were defined according to the position of the local free energy minima identified in CV space. Core sets were considered to identify state-to-state transitions as reported in Fig. ~\ref{panel1}a). The equilibrium probability obtained from the MSM is in good agreement with the result obtained from WTmetaD (Supplementary Material, Fig. 6), providing a consistency check for our sampling strategy. Relaxation times were derived for each solvent from the MSM model and are reported in \figurename{~\ref{fig:solution_metad}}b). Relaxation times show that, while the equilibrium distribution of conformers varies only slightly between solvents, the system equilibration dynamics differs more substantially. The slowest equilibration in solution is computed in water, where the system equilibrates in just under 1 ns. The second longest time is registered for 1-butanol. In the remaining solvents relaxation times are all of the order of half a nanosecond. These results indicate that hydrogen bonding with the solvent are able to affect the interconversion dynamics of ibuprofen conformational isomers in solution, even if their contribution to the equilibrium distribution of conformers is limited. 

\subsection{Equilibrium distribution and lifetime of ibuprofen conformers in the crystal bulk}

WTmetaD was also employed to sample the conformational space accessible to an ibuprofen molecule in the bulk of the crystal. The free energy surface computed for this case is reported in Fig. \ref{fig:thermodynamics}b). An analysis of the associated conformer equilibrium distribution revealed that, while conformer c1 has a 98 \% probability, ibuprofen maintains a non-negligible conformational flexibility in the crystal bulk. Conformers involving a rotation along the local torsional angle can be observed and account for 2\% of the equilibrium distribution. 
With the aid of \emph{infrequent} metadynamics, the lifetime of conformer c1 was estimated to be of the order of  100 ns. The p-value associated with the Kolmogorov-Smirnov statistic of the escape time distribution\cite{Salvalaglio2014} is reported in Tab. VII in the Supplementary Material. In all of the trajectories used to compute the escape time from the local minimum corresponding to conformer c1, the escape leads to conformer c2.

\subsection{Equilibrium distribution and lifetime of ibuprofen conformers at the crystal-solution interface}

In this section we report the results obtained for an ibuprofen molecule, adsorbed on the crystal surface (state b in Fig. \ref{fig:states_definitions}), or embedded in the surface layer in contact with the liquid solution  (state c in Fig. \ref{fig:states_definitions}). We find that the conformational population and the system equilibration times in the states at the crystal solution/interface depend on the structure of the surface, as well as on the polarity of the functional groups in contact with the solution (see \figurename{~\ref{fig:surface_layers}} for polar and apolar layer descriptions). To quantify the effect of the solvent on the surface roughness and stability we use the order parameter SMAC. A detailed overview of the effect of each solvent on the behaviour of the both apolar and polar \{100\} crystal faces can be found in Section C of the Supplementary Material. In Tab. ~\ref{tab:states_solvent} we summarise the effect of surface-solvent interactions on the surface roughness and stability of the adsorbed state.

	\begin{table}[!h]
	\caption{Interface structure and stability of the adsorbed state for all interface-solvent combinations.}
	\begin{tabular}{c|c|c|c|c}
		\hline 
		\textbf{solvent} & \multicolumn{4}{c}{\textbf{interface}} \\ 
		&\multicolumn{2}{c|}{polar} & \multicolumn{2}{c}{apolar} \\ 
		& structure & ads. state & structure & ads. state \\ 
		\hline 
		water & ordered & no & ordered & yes\\

		1-butanol & disordered & yes & ordered & yes  \\

		toluene & unstable & - & ordered & no\\

		cyclohexanone & ordered & yes & ordered & yes \\ 

		cyclohexane & unstable & - & ordered & yes\\

		acetonitrile & disordered & yes & ordered & no \\

		trichloromethane & unstable & - & ordered & no \\
		\hline 
	\end{tabular} 
	\label{tab:states_solvent}
\end{table}

\begin{figure*}[h!tpb]
	\includegraphics[width=\linewidth]{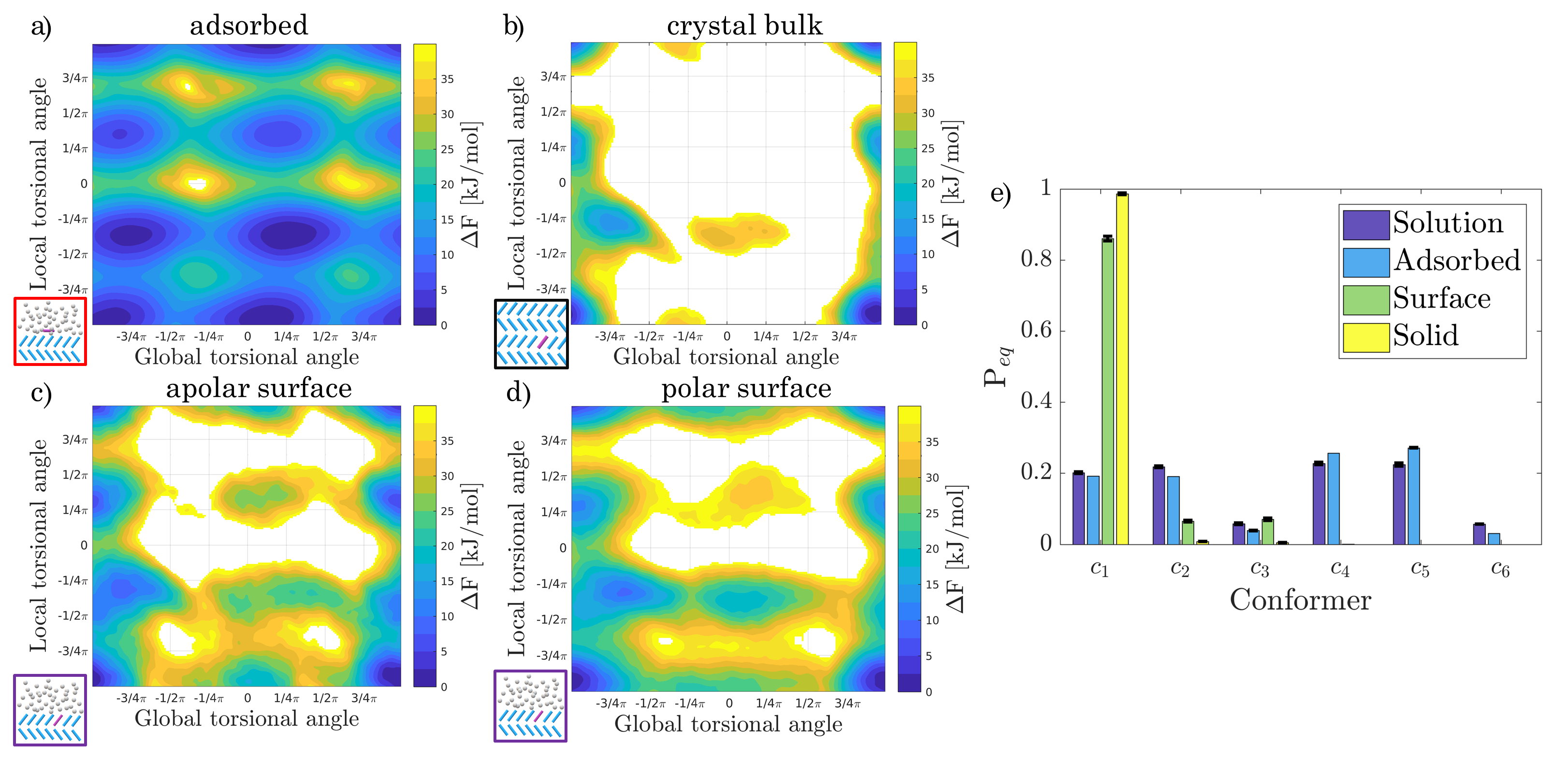}
	\centering
	\caption{a-d) Free energy surfaces of an ibuprofen molecule in water, adsorbed on the solid/liquid interface (a), in the crystal bulk (b), embedded in the apolar crystal surface (c) and in the polar crystal surface (d). It can be noted that the accessible conformational space is systematically reduced as the molecule moves from solution to the crystal bulk and that the topology of the surface changes moving from the adsorbed to the surface state. e) Equilibrium probability of the six ibuprofen conformers c1 - c6, found in the stages involving incorporation of a molecule into the crystal, considering water as a solvent and an apolar \{100\} crystal surface. }
	\label{fig:thermodynamics}
\end{figure*}

\begin{figure}[h!tpb]
	\includegraphics[width=\linewidth]{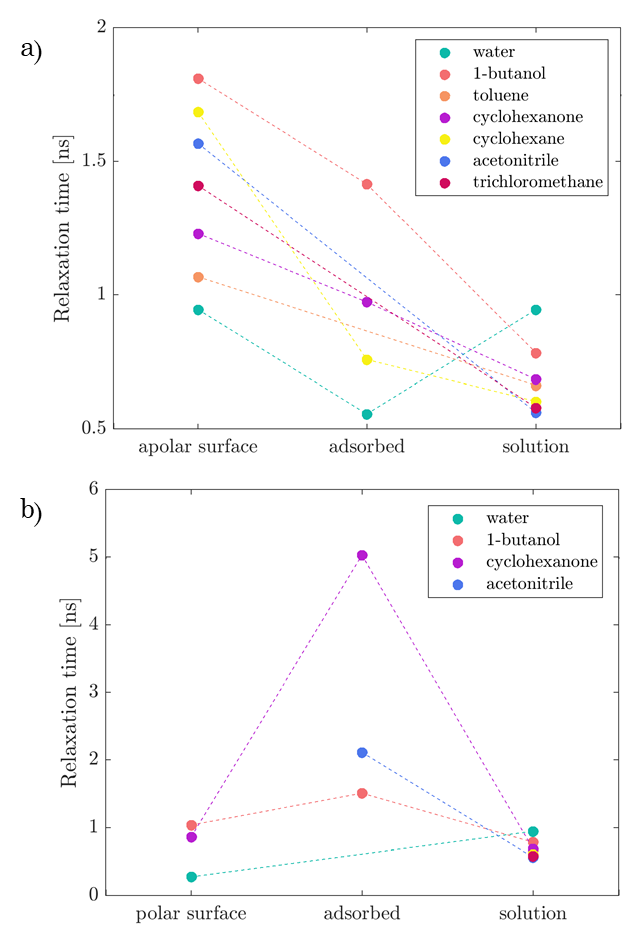}
	\centering
	\caption{System relaxation times, obtained through an MSM analysis in the case of a) an apolar \{100\} surface and b) polar \{100\} surface.  }
	\label{fig:relaxation_times}
\end{figure}

Free energy surfaces for an ibuprofen adsorbed on the crystal surface and embedded in the outer layer of the crystal configurations for the water/ibuprofen interface are reported in Fig. \ref{fig:thermodynamics} a), c) and d).  Equilibrium probability of the c1 - c6 conformers moving from the bulk of the liquid (water) to the bulk of the crystal are reported in Fig. \ref{fig:thermodynamics}e). Characteristic relaxation times of the conformer population for surface/adsorbed and solution states for the apolar and polar surface layers are reported in Fig. \ref{fig:relaxation_times}a) and in Fig. \ref{fig:relaxation_times}b) respectively . 

\subsubsection*{Ibuprofen adsorbed on the crystal surface}
The equilibrium distribution of ibuprofen conformers adsorbed on the \{100\} crystal surface resembles closely the distribution obtained in solution. Slightly higher probability of c2, c4 and c5 is observed in certain solvents (see section F of the Supplementary Material), revealing signs of conformationally-selective adsorption.
MSMs were built using unbiased MD simulations for the stable adsorbed states. The model was again verified by comparing the equilibrium distribution from MSM to the one recovered from WTmetaD (see Fig. 7 in Supplementary Material). 
The interplay between solvent and interface, however, has in this case a much more remarkable effect on the system dynamics.   Relaxation times in the adsorbed state on an apolar surface show a different trend with respect to those computed in solution. For instance, in the adsorbed state equilibration is the slowest in 1-butanol, 1.4 ns, followed by cyclohexanone of around 1 ns. 
Contrary to what is observed in the solution bulk, conformational relaxation in the adsorbed state is the fastest for the crystal/water interface.
Equilibration time on the polar surface for the case of 1-butanol is the same as in the adsorbed configuration on the apolar surface. In the case of cyclohexanone, however, equilibration on the polar surface is significantly slower, 5 ns, than its counterpart on the apolar surface. This five-fold increase in the relaxation time shows that the character  of the surface impacts dynamics more significantly than thermodynamics, i.e. differences in the equilibrium probability of conformers in the two surfaces in fact do not exceed 10\%. A complete summary of conformational relaxation times for the adsorbed state can be found in Tab. ~\ref{relaxation_solvent}.
These results clearly indicate that the kinetics of conformational rearrangement of adsorbed molecules are dictated by specific solvent-surface interactions, and their relaxation timescale differs from that computed in solution.

\begin{table}[!h]
\caption{Relaxation times [ns] for the adsorbed states on polar and apolar \{100\} ibuprofen surface. Only cases in which the adsorbed state is stable have been reported.}
\begin{tabular}{c|c|c}
\hline 
\textbf{solvent} & \multicolumn{2}{c}{\textbf{Relaxation times [ns]}} \\ 
&{polar} & {apolar} \\ 
\hline 
water &  - & 0.6\\
1-butanol & 1.5 & 1.4  \\
toluene &  - & - \\
cyclohexanone &  5.0 & 0.9 \\ 
cyclohexane & - &  0.8 \\
acetonitrile & - & 2.1 \\
trichloromethane & - & -  \\
\hline 
\end{tabular} 
\label{relaxation_solvent}
\end{table}

\subsection*{Ibuprofen embedded in the crystal surface}

The equilibrium distribution within the surface layer of the \{100\} crystal is dominated by the c1 conformer with an equilibrium probability higher than 80\%, for both the apolar and polar faces. The probability distribution suggests that rotations along the \emph{local} torsional angle are associated to a lower energy penalty regardless of whether the CH$_3$ groups are in direct contact with the solvent. Particularly low probability for c1 is found for the case of 1-butanol in the polar surface. This is due to surface roughening, which allows for a higher conformational flexibility. (Fig. 5 in Section F of the Supplementary Material)
In the case of the polar surface in acetonitrile we find that perturbing the conformational space of a surface molecule, by adding biasing potential within the framework of WTmetaD, causes its detachment, indicating that  detachment from the surface is coupled with its conformational transition. 
Overall, highlighting equilibrium distribution variations between states involved in dissolution shows that embedding an ibuprofen molecule in the surface layer provides a significant conformational selectivity. In fact, c1 $\rightarrow$ c4 and c1 $\rightarrow$ c5 become so rare, that the probability of c4 and c5 reduces to less than 1 \%.
We build MSMs for all surface/solvent couples, based on the kinetics of state-to-state transitions, obtained from WTmetaD simulations. The equilibrium distribution obtained is once again in good agreement with the one from WTmetaD (Fig. 8 in the Supplementary Material), confirming the internal consistency of the results obtained.   With the exception of water, relaxation times computed from MSMs show that conformational equilibration in the apolar surface increase substantially for all solvents compared to the solution case.  Relaxation times range between 1 and 2 ns (2-5 times slower than in solution, see Fig. \ref{fig:relaxation_times}a), with the largest slowdown registered for the case of 1-butanol. Conformational equilibration of ibuprofen in the polar \{100\} surface on the other hand, is characterised by the same relaxation time as in solution (see Fig. \ref{fig:relaxation_times}b), around 1 ns. Water represents once again an exception. We observe that its relaxation time in the polar surface is counter-intuitively \emph{faster} than in the bulk solution (0.2 vs 1 ns).

\begin{figure*}[h!tpb]
	\includegraphics[width=0.9\linewidth]{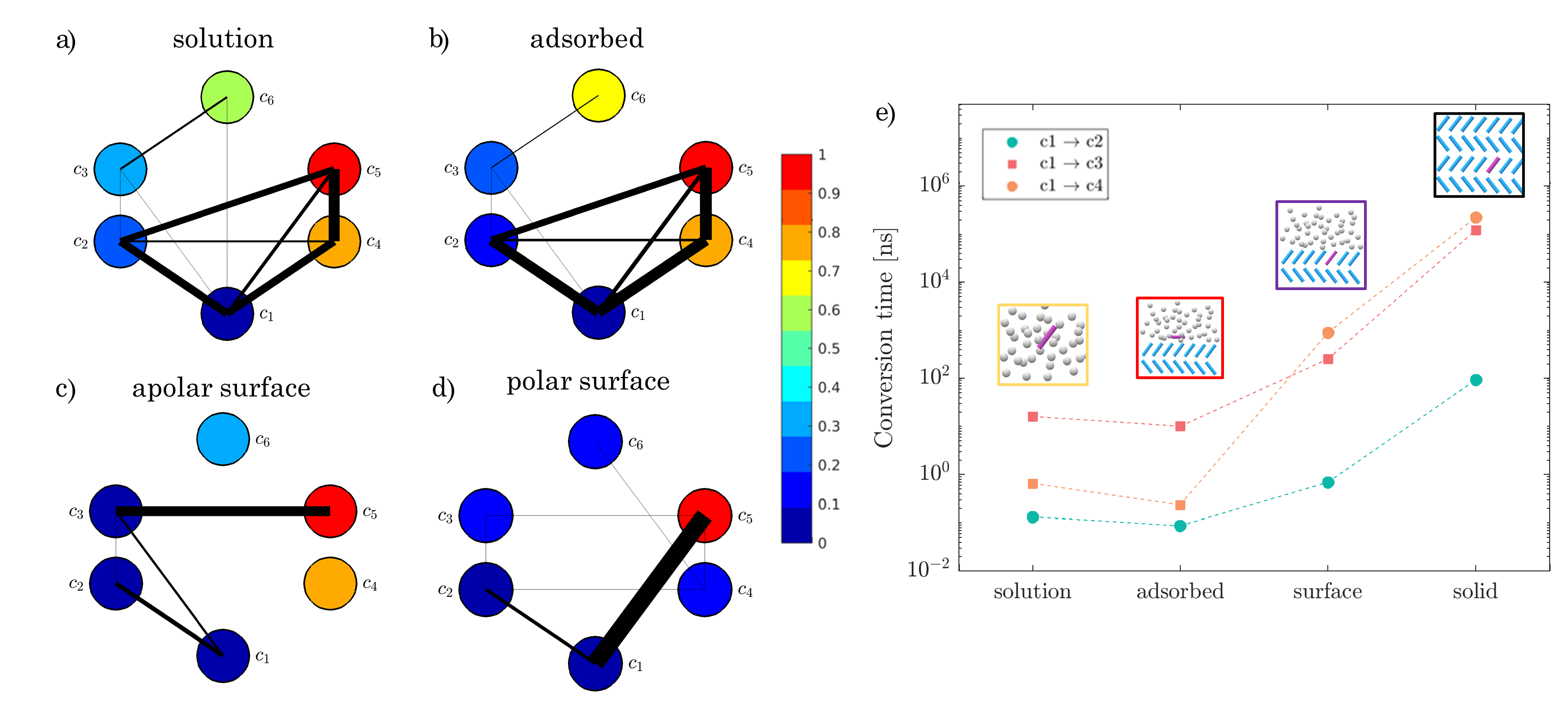}
	\centering
	\caption{a) Mechanism of conversion of conformer c1 to conformer c5, in solution (a), an adsorbed configuration on the apolar \{100\} crystal surface (b), embedded in the apolar \{100\} crystal surface (c) and in the polar \{100\} crystal surface (d). In all four cases the solvent is water. The colour of the states represents the probability of converting to conformer c5. The net probability flux is represented through line thickness. b) Conversion rates between c1 and neighbouring conformers in CV space going through the four states considered in this study. Data points marked with circles have been obtained through unbiased MD or recovered from well-tempered metadynamics, while the data points in squares have been computed using known transition rates and free energy barriers between relevant states.}
	\label{relaxation_times}
\end{figure*}

\subsection{Discussion}

\paragraph*{Conformational equilibrium distribution}
In this study the conformational flexibility of ibuprofen is investigated with the aid of MD simulations in combination with WTmetaD and the theoretical framework of the Markov State Model. In all cases we find good agreement between the equilibrium distribution obtained from WTmetaD and the MSM constructed from individual transition rates. Based on these findings, we can conclude that the free energy landscape of ibuprofen conformers is strongly dependent on the environment (solution, adsorbed, interface, crystal bulk) and weakly dependent on the solvent. The latter is further confirmed by the equilibrium distribution obtained with DFT/COSMO calculations, which also predict minimal variability in the equilibrium distribution of conformers in different solvents.  Conformational selectivity is observed in the adsorbed state and the surface state of the ibuprofen molecule, and is particularly obvious in the crystal bulk. Despite the fact that the polar and apolar \{100\} surfaces expose different functional groups of the molecule to the solution, we find that the conformational distribution of ibuprofen in both is very similar, indicating that rotation along the local torsional angle is more energetically favourable than along the global one in both cases.

\paragraph*{Relaxation dynamics of the conformer population} We use MSMs to obtain the characteristic equilibration time for the population of conformers c1 - c6 in all of the states investigated. In most cases the presence of a surface only marginally slows down the global conformational rearrangement. Nonetheless, it can be concluded that the system relaxation times are both environment and solvent dependent. More importantly, there is a significant difference in the computed trends between the polar and apolar \{100\} surfaces. Whilst in the apolar \{100\} surface case the system equilibration generally increases when moving from solution to the surface, in the case of the polar surface the relaxation times increase in the adsorbed state and decrease again in the surface.  The solid/liquid interface of the \emph{polar} and \emph{apolar} interfaces has significantly different roughness. We can conclude that conformational rearrangement of ibuprofen in both the surface and in the adsorbed state depends strongly on specific solvent-surface interactions. 

\paragraph*{Equilibration mechanism} 

While the global relaxation time of the conformational population slows down only marginally, there is a clear modification in the mechanism of conformational rearrangement of ibuprofen molecules when moving from considering the bulk solution to the crystal surface. 
To quantify this change in terms of kinetics, we investigate the escape rates from the bulk crystal conformer c1 to each of its neighbours in CV space, c1 $\rightarrow$ c2, c1 $\rightarrow$ c3 and c1 $\rightarrow$ c4, in all stages of incorporation with respect to the apolar \{100\} crystal slab. Transition rates that have been obtained with the aid of metadynamics and/or the analysis of unbiased MD trajectories are reported with circles in ~\figurename{~\ref{relaxation_times}e)}. The rest of the transition rates have been calculated using the ratio of free energy barriers between states from WTMetaD calculations (indicated with squares).  
We find that the conversion c1 $\rightarrow$ c4 goes from being comparable to the c1 $\rightarrow$ c2 conversion in solution to being the slowest transition in the crystal surface and in bulk.  This indicates that the rearrangement mechanism of conformers depends heavily on the position with respect to the crystal interface. 

In order to further investigate modifications in the conformational transition mechanism, we use the MSMs to analyse a transition pathway from conformer c1 to conformer c5. We find that in solution and in the adsorbed state the most likely mechanism of conversion unfolds via conformers c2 and c4, as indicated by the net flux in \figurename{~\ref{relaxation_times}}a) and b). Looking at the mechanism of conversion in the surface however, we find that it is dependant on the surface-solvent interactions. For the case of an ibuprofen in the apolar surface c1 to c5 transition occurs exclusively via conformer c3 \figurename{~\ref{relaxation_times}}c), whilst for the case of a polar surface the most likely mechanism is a direct transition \figurename{~\ref{relaxation_times}}d).

\section{\label{sec:conslusions}Conclusions}
By combining extensive unbiased MD, WTMetaD and MSMs, in this study we propose a systematic approach at the identification of kinetics, thermodynamics and transition mechanisms of relevant conformational transitions of organic molecules at the crystal/solution interface.

Here we investigate conformational isomerism of ibuprofen at the solid-liquid interface. We find that the equilibrium distribution of conformers is strongly dependent on the environment of the molecule with respect to the crystal surface. When in contact with the \{100\} surface, the conformational rearrangement time generally increases  moving from the solution to the surface states, and differs between solvents. On the other hand, the conformational rearrangement in the polar \{100\} surface exhibits different kinetic behaviour and relaxation times in the surface are very similar to those in solution. We find that the specific solvent/surface interactions govern the conformational behaviour of the molecule. Moreover, in both the polar and apolar surfaces we find a change in the mechanism of conformational rearrangement with respect to bulk solution. 
With these results we show that, even in a moderately flexible molecule such as ibuprofen, thermodynamics, kinetics and mechanisms of conformational equilibria are significantly affected by the presence of a crystal interface and by its interplay with the solvent.  
This finding highlights the need to develop systematic approaches for including conformational transitions in mesoscopic models of crystal growth.

\section*{Conflict of interest}
There are no conflicts to declare.

\section*{\label{sec:acknowledgments}Acknowledgements}
The        authors        acknowledge       Legion        High        Performance        Computing        Facility                for        access        to        Legion@UCL        and        associated        support        services,        in        the       completion       of       this       work.\\

\section*{Supplementary Material}
See supplementary material for a list of simulations, the set up of the collective variables used to characterise the surface structures, a detailed description of the surface-solution interactions observed, free energy surfaces for all the states in all solvents considered, convergence of free energy differences, and conformer populations in all states computed with WTmetaD and MSM, and conformers probability in solution computed with the DFT/COSMO procedure.

\section*{\label{sec:references}References}
\bibliographystyle{unsrt}

\newpage 

$ $

\newpage

\onecolumngrid
\begin{center}
  \textbf{\large Dynamics and thermodynamics of Ibuprofen conformational isomerism at the crystal/solution interface. \\Supplementary Material}\\[.2cm]
Veselina Marinova$^1$, Geoffrey P. F. Wood$^2$, Ivan Marziano$^3$, Matteo Salvalaglio$^{*1}$ \\ 
$^1$Thomas Young Centre and Department of Chemical Engineering, University College London, London WC1E 7JE, UK. \\
$^2$Pfizer Worldwide Research and Development, Groton Laboratories, Groton, Connecticut 06340, USA \\%
$^3${Pfizer Worldwide Research and Development, Sandwich, Kent CT13 9NJ, UK} \\%
$^*$m.salvalaglio@ucl.ac.uk

(Dated: \today)\\[1cm]
\end{center}
\twocolumngrid

\setcounter{equation}{0}
\setcounter{figure}{0}
\setcounter{table}{0}
\setcounter{page}{1}
\renewcommand{\theequation}{S\arabic{equation}}
\renewcommand{\thefigure}{S\arabic{figure}}
\renewcommand{\bibnumfmt}[1]{[S#1]}
\renewcommand{\citenumfont}[1]{S#1}

\date{\today}

\section*{Supplementary Material}
\subsection{Simulations performed}
Tab. \ref{simulations} provides a summary of all simulations performed towards the characterisation of the thermodynamics and the kinetics of the conformational transitions of ibuprofen. In Tab. ~\ref{computational_time} the approximate computational time [ns] taken for each type of simulation has been summarised.
\paragraph{Bulk Solution} In bulk solution, the information on the equilibrium probability of ibuprofen conformers was extracted from MD with WTmetaD simulations, one for each of the solvents. Kinetic information was obtained by analysing one unbiased MD trajectory per solvent and recording the state-to-state transitions.
\paragraph{Adsorbed state} In the adsorbed states, thermodynamic information was, again, extracted with the aid of WTmetaD, however, additional static bias was applied to prevent detachment from the surface and increase computational efficiency. Kinetic transitions between states were extracted from unbiased MD simulations for those solvents which stabilise an adsorbed state. Fifty MD simulations were performed for each solvent case to ensure thorough sampling as static bias was not applied in this case and simulations were stopped upon detachment of the molecule from the surface. The duration of the simulations was therefore the lifetime of the state in the range of a few nanoseconds.
\paragraph{Surface state}In the crystal surface, similarly to the other cases, thermodynamic information was obtained with the aid of WTmetaD for each solvent case. Kinetic information was recovered from WTmetaD simulations for computational efficiency. To this aim, all six conformers were categorised as starting configurations and 50 simulations per starting configuration were performed. The simulations were stopped upon observing a conformational transition. This procedure was carried out for all solvents where a stable surface state exists.
\paragraph{Crystal bulk} Finally, a single MD with WTmetaD simulation was carried out to obtain the conformational equilibrium probability of ibuprofen in the crystal bulk. In the crystal, kinetic information was obtained only for conformer c1 as a starting configuration with the aid of WTmetaD. 30 simulations were carried out for the purpose.

	\begin{table}
		
		\caption{Summary of simulations carried out in the conformational study of ibuprofen}	
		\begin{tabular}{ |c|c  c|c  c|} 
			\hline
			state  	 & \multicolumn{2}{c|}{\textbf{thermodynamics}} & \multicolumn{2}{c|}{\textbf{kinetics}}\\
			\hline
	   bulk solution & \textbf{1}$\times$   &  water           &       \textbf{1}$\times$      &  water             \\
					 &   		  		    &  1-butanol       &             		       &  1-butanol         \\
					 &			  			&  toluene         &               		   &  toluene           \\
			         &            			&  cyclohexanone   &               		   &  cyclohexanone     \\
			         &            			&  cyclohexane     &          		       &  cyclohexane       \\
			         &            			&  acetonitrile    &               		   &  acetonitrile      \\
			         &            			&  trichloromethane&    				       &  trichloromethane  \\
			
			\hline
		    adsorbed & \textbf{1}$\times$   &  water           & \textbf{50}$\times$     &  water \\
			state on & 			  		    &  1-butanol       &           		       &  1-butanol \\
   an \textbf{apolar}&			  			&  toluene         &               		   &  cyclohexanone \\
	  \{100\} surface&            			&  cyclohexanone   &             		       &  cyclohexane \\
					 &            			&  cyclohexane     &       		           &   \\
					 &            			&  acetonitrile    &                		   &   \\
					 &            			&  trichloromethane&     				       &  \\
			
			\hline
			adsorbed & \textbf{1}$\times$   &  water              &\textbf{5}$\times$     &  1-butanol \\
			state on & 			  		    &  1-butanol          &           		       &  cyclohexanone \\
	 a \textbf{polar}&			  			&  toluene            &             		   &  acetonitrile \\
	  \{100\} surface&            			&  cyclohexanone      &            		       &  \\
			 		 &            			&  cyclohexane        &        		           &   \\
					 &            			&  acetonitrile       &             		   &   \\
					 &            			&  trichloromethane  &  				       &   \\
		
			\hline
			 surface & \textbf{1}$\times$   &  water           &    \textbf{6}$\times$\textbf{50}$\times$      &  water \\
			state in & 			  		    &  1-butanol       &             		       			   		   &  1-butanol\\
   an \textbf{apolar}&			  			&  toluene         &               		   			   		   &  toluene \\
	  \{100\} surface&            			&  cyclohexanone   &               		       			   		   &  cyclohexanone\\
				   	 &            			&  cyclohexane     &           		           			   		   &  cyclohexane \\
					 &            			&  acetonitrile    &                		   			   		   &  acetonitrile\\
					 &            			&  trichloromethane&     				       			   		   &  trichloromethane \\
			\hline
		    surface & \textbf{1}$\times$    &  water              &\textbf{6}$\times$\textbf{50}$\times$    &  water \\
		   state in & 			  		    &  1-butanol          &           		       			   		   &  1-butanol\\
    a \textbf{polar}&			  			&  toluene            &             		   			   		   &  cyclohexanone \\
     \{100\} surface&            			&  cyclohexanone      &            		       			   		   &  \\
				    &            			&  cyclohexane        &        		           			   		   &   \\
		    		&            			&  acetonitrile       &             		   			   		   &   \\
				    &            			&  trichloromethane   &  				       			   		   &   \\
	
		   \hline
     crystal bulk   & \textbf{1}$\times$    &  no solvent         &\textbf{1}$\times$\textbf{30}$\times$    &  no solvent \\
			\hline
	\end{tabular}
		\label{simulations}
		
	\end{table}

\begin{table*}[h!]
	\caption{Table of approximate computational time [ns] for each type of simulation performed}	
	\begin{tabular}{ |c|c|c|c|c|c|c|} 
		\hline
		& \textbf{bulk}  & \multicolumn{2}{c|}{\textbf{adsorbed state}}  & \multicolumn{2}{c|}{\textbf{surface state}}& \textbf{crystal}\\
		& \textbf{solution}  & apolar & polar & apolar & polar&\textbf{bulk} \\
		\hline
		thermodynamics  & 150  & 100 & 	300  & 300 & 150 & 45 \\
		kinetics	    & 250  & 2-5 & 2-5 & up to 2& up to 2 &2-3\\
		
		\hline
	\end{tabular}
	\label{computational_time}

\end{table*}

 \begin{figure*}
	\includegraphics[width=0.8\linewidth]{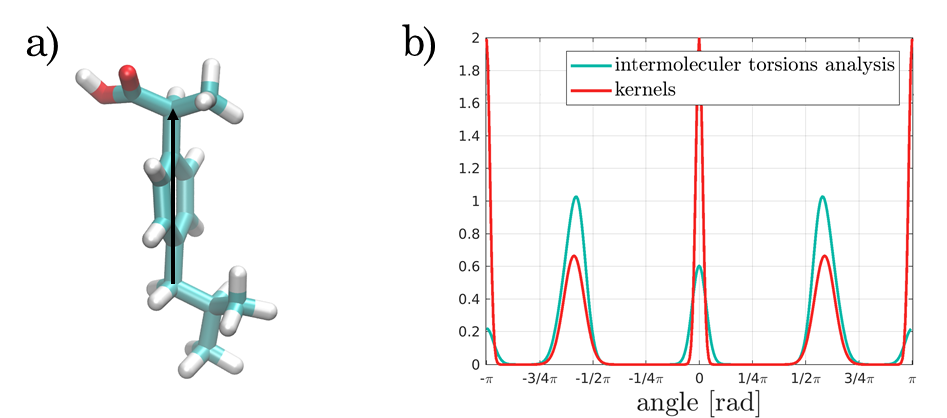}
	\centering
	\caption{a) Figure shows the vector description of the ibuprofen molecule in the crystal bulk. b) The plot in green shows the characteristic intermolecular torsional angles distribution between molecular vectors in the crystal bulk, obtained with the INTERMOLECULARTORSIONS utility in Plumed 2.3. The plot in red shows the Gaussian function kernels set up in SMAC. }
	\label{smac_setup}
	
\end{figure*}

\subsection{SMAC parameters} The collective variable SMAC, as implemented in Plumed 2.3, was used to quantify the roughness of the surface/solution interface. The implementation of the variable involves representing molecules of interest in a vector form, as shown in \figurename{~\ref{smac_setup}}a). The intermolecular torsional angles distribution, characteristic for the crystal bulk arrangement, is then studied to set up the variable. The intermolecular torsions between the vectors defined for ibuprofen were analysed in the crystal bulk and their characteristic fingerprint, shown in \figurename{~\ref{smac_setup}}b) in green, was found. The SMAC variable requires an input of this characteristic angle distribution in the form of Gaussian functions kernels, which are used as a reference in the evaluation of the degree of order of a cluster of molecules. The kernels used for the setup are shown in \figurename{~\ref{smac_setup}}b) in red. The plumed file input for the SMAC variable is also provided below.
\begin{flushleft}
\noindent
SMAC ... \\
SPECIES=molecules \\
SWITCH=\{RATIONAL R\_0=0.7\} \\
MEAN \\
KERNEL1=\{GAUSSIAN CENTER=-3.14 SIGMA=0.1\} \\
KERNEL2=\{GAUSSIAN CENTER=-1.85 SIGMA=0.15\} \\
KERNEL3=\{GAUSSIAN CENTER=0.0 SIGMA=0.1\} \\
KERNEL4=\{GAUSSIAN CENTER=1.85 SIGMA=0.15\} \\
KERNEL5=\{GAUSSIAN CENTER=3.14 SIGMA=0.1\} \\
SWITCH\_COORD=\{RATIONAL R\_0=3.0\} \\
LABEL=smac \\
... SMAC \\
\end{flushleft}

\begin{figure}
\includegraphics[width=0.99\linewidth]{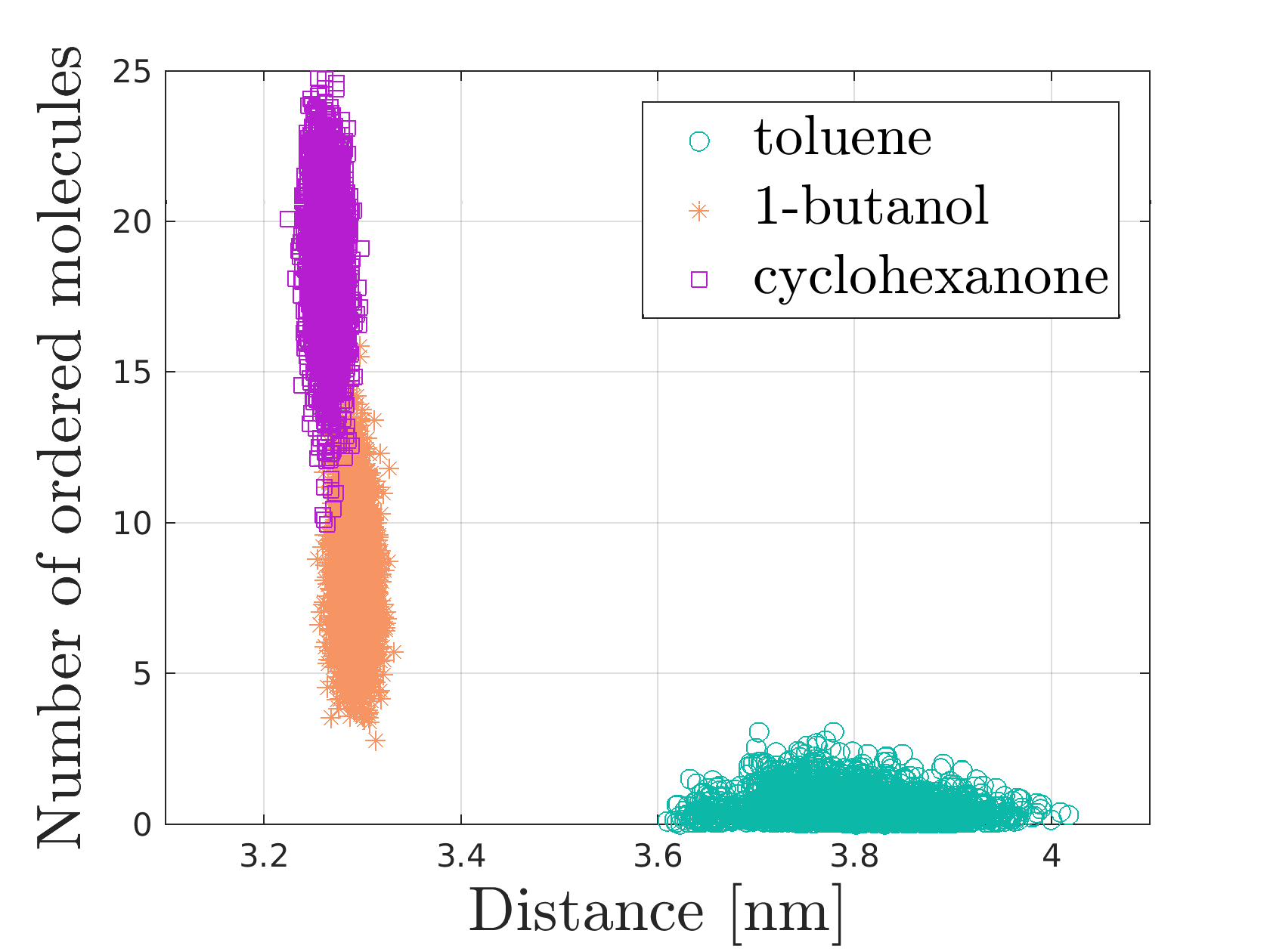}
	\centering
	\caption{Plot of number of ordered molecules at the solution/surface interfaceof the polar \{100\} ibuprofen face vs. distance of the surface layer from the pseudobulk of the crystal slab. State of the surface is analysed after 1-3 ns from the begining of the simulation to ensure that equilibrated configuration has been reached. The plot shows three characteristic cases of the solvent effect on the structure of the surface. In the case of toluene (green), full dissolution of the surface is observed, hence the surface is categorised as unstable. For the case of cyclohexanone (purple), the surface remains stable and ordered. In the case of 1-butanol (orange) as a solvent, the crystal surface is stable, but roughened. }
	\label{fig:surface_description}
	
\end{figure}

\subsection{Analysis of the interface structure}

After addressing the thermodynamics and kinetics of the conformational transitions of an ibuprofen molecule in bulk solution, we report results for the \{100\} crystal interface. As briefly mentioned in the Methods section we note that, due to the layered nature of the ibuprofen crystal, the \{100\} face exposes either a \emph{polar} or an \emph{apolar} surface to solution. The polar \{100\} face is characterised by the exposure of carboxylic groups  towards the solution. The apolar \{100\} crystal face, on the other hand, exposes methyl groups. To analyse structural features of the polar and apolar layers at the crystal/solution interface we carry out unbiased MD simulations. The degree of order of polar and apolar crystal interfaces is monitored during the relaxation dynamics through the calculation of the SMAC order parameter we referred to in the methods, developed ad-hoc for ibuprofen (for the details see Section B). The dissolution of molecules belonging to the surface layer was monitored by computing their distance from the bulk. In Figure \figurename{~\ref{fig:surface_description}} we report the configurations sampled by three different  surface/solution models in the order parameter / distance space.  Three different limiting behaviours emerge: stable and smooth interface (purple set), stable and disordered interface  (orange set),  and unstable interface that spontaneously dissolves (green set). 

	\begin{table*}[!h]
	\caption{Interface structure and stability of the adsorbed state for all interface-solvent combinations.}
	\begin{tabular}{c|c|c|c|c}
		\hline 
		\textbf{solvent} & \multicolumn{4}{c}{\textbf{interface}} \\ 
		&\multicolumn{2}{c|}{polar} & \multicolumn{2}{c}{apolar} \\ 
		& structure & ads. state & structure & ads. state \\ 
		\hline 
		water & ordered & no & ordered & yes\\
		
		1-butanol & disordered & yes & ordered & yes  \\
		
		toluene & unstable & - & ordered & no\\
		
		cyclohexanone & ordered & yes & ordered & yes \\ 
		
		cyclohexane & unstable & - & ordered & yes\\
		
		acetonitrile & disordered & yes & ordered & no \\
		
		trichloromethane & unstable & - & ordered & no \\
		\hline 
	\end{tabular} 
	\label{states_solvent}
\end{table*}

	The apolar \{100\} face is stable and ordered in all of the solvents considered and does not show signs of roughening within simulation times of a few hundred nanoseconds. On the other hand, the polar \{100\} face displays a much more dynamic behaviour, strongly influenced by the solvent: while certain solvents cause roughening of the surface or its complete dissolution, others promote the stabilization of an ordered interface. 
	Furthermore, we observe that an ibuprofen molecule adsorbed on the crystal surface is not stable for all the interface (polar/apolar) solvent combinations. In particular, in the case of an apolar surface, the adsorbed state exists for the case of water, 1-butanol, cyclohexanone and cyclohexane. The polar \{100\} surface was found to dissolve in toluene, cyclohexane and trichloromethane. This suggests that low polarity solvents favour an interaction between the carboxylic groups of neighboring ibuprofen molecules, promoting the distortion and dissolution of the surface. An adsorbed state in the case of the polar surface was found to be stable only in 1-butanol, cyclohexanone and acetonitrile. A summary of the interface structural analysis is reported in Tab. ~\ref{states_solvent}. In the  analysis of the ibuprofen conformational space and transition dynamics at the crystal/solution interface we exclude all unstable states uncovered in the analysis of the interface structure.

\subsection{System relaxation time}
System relaxation time was obtained by setting up a Markov State Model, where each ibuprofen conformer is considered a microstate. Tab. ~\ref{relax} summarises conformational equilibration times of an ibuprofen molecule for all states considered. 

	\begin{table*}[h!]
	\caption{Table of system relaxation time [ns], recovered for a MSM, for each molecule state considered in the study.}	
	\begin{tabular}{ |c|c|c|c|c|c|} 
	\hline
					  	 & \textbf{bulk}  & \multicolumn{2}{c|}{\textbf{adsorbed state}}  & \multicolumn{2}{c|}{\textbf{surface state}}\\
              		     & \textbf{solution}  & apolar & polar & apolar & polar \\
		\hline
		 			   water  			& 0.94  & 0.57 & -	  & 0.94 & 0.27 \\
					   1-butanol	    & 0.78 	& 1.58 & 1.51 & 1.81 & 1.04 \\
				   	   toluene		 	& 0.66 	&  -   &  -   & 1.07 & -    \\
					   cyclohexanone	& 0.68 	& 1.08 & 5.03 & 1.23 & 0.86 \\
					   cyclohexane  	& 0.60 	& 0.79 &  -   & 1.68 & -    \\
					   acetonitrile 	& 0.56 	& -    & 2.11 & 1.57 & -    \\
					   trichloromethane & 0.58 	& 	-  & -    & 1.41 & -    \\
					   
	  \hline
	\end{tabular}
	\label{relax}

\end{table*}

\begin{table*}[h!]
	\caption{Table of residence time [ps] of each conformer for the case of an ibuprofen molecule in bulk solvent.}	
	\begin{tabular}{ |c|c|c|c|c|c|c|} 
		\hline	
		\textbf{bulk solution}		 & \textbf{c1} & \textbf{c2} & \textbf{c3} & \textbf{c4} & \textbf{c5} & \textbf{c6} \\
		& [ps]  & [ps] & [ps] & [ps] & [ps] & [ps]  \\
		\hline
		
		water & 102 & 107 & 675 & 112& 110 & 485 \\
		
		1-butanol & 140 & 129 & 671 & 136 & 143 & 512 \\
		
		toluene & 101 & 122 & 390 & 114 & 104 & 420 \\
		
		cyclohexanone & 162 & 164 & 467 & 156 & 154 & 481 \\
		
		cyclohexane & 134 & 158 & 366 & 161 & 147& 325 \\
		
		acetonitrile & 94 & 105& 305 & 113& 105& 428 \\
		
		trichloromethane & 89 & 90 & 268 &90 &96 & 307 \\
		\hline
	\end{tabular}
	\label{solution_lifetimes}
	
\end{table*}

\begin{table*}[h!]
	\caption{Table of residence time [ps] of each conformer for the case of an ibuprofen molecule adsorbed on the crystal surface.}	
	\begin{tabular}{ |c|c|c|c|c|c|c|} 
		\hline
		\textbf{adsorbed on an} & \textbf{c1} & \textbf{c2} & \textbf{c3} & \textbf{c4} & \textbf{c5} & \textbf{c6} \\
		\textbf{apolar surface} & [ps]  & [ps] & [ps] & [ps] & [ps] & [ps]  \\
		\hline
		
		water & 95 & 85 & 381 & 91 & 92 & 429 \\
		
		1-butanol & 166 & 143 & 623 & 148 & 157 & 162 \\
		
		cyclohexanone & 168 & 179 & 675 & 152 & 146 & 1053 \\
		
		cyclohexane & 155 & 148 & 579 & 140 & 183& 478 \\
		\hline
		\textbf{adsorbed on a} & \textbf{c1} & \textbf{c2} & \textbf{c3} & \textbf{c4} & \textbf{c5} & \textbf{c6} \\
		\textbf{polar surface} & [ps]  & [ps] & [ps] & [ps] & [ps] & [ps]  \\
		\hline
		
		1-butanol & 149 & 197 & 1316 & 169 & 138 & 123 \\
		
		cyclohexanone & 127 & 180 & 1187 & 174 & 145 & 5283 \\
		
		acetonitrile & 147 & 168 & 1998 & 179 & 213 & 603 \\
		\hline
	\end{tabular}
	\label{adsorbed_lifetimes}
\end{table*}

\begin{table*}
	\caption{Table of residence time [ps] of each conformer for the case of an ibuprofen molecule in the crystal surface and in the crystal bulk. The residence time have been recovered from WTmetaD simulations and include the KS statistics p-values for a 0.95 significance rate.}	
	\begin{tabular}{ |c|cc|cc|cc|cc|cc|cc|} 
		\hline
		\textbf{state in the \{100\}} & \multicolumn{2}{c|}{\textbf{c1}} & \multicolumn{2}{c|}{\textbf{c2}} &\multicolumn{2}{c|}{\textbf{c3}} & \multicolumn{2}{c|}{\textbf{c4}} & \multicolumn{2}{c|}{\textbf{c5}} & \multicolumn{2}{c|}{\textbf{c6}}  \\
		\textbf{apolar surface} & [ps] & p-value & [ps] & p-value &[ps] & p-value &[ps] & p-value &[ps] &p-value & [ps]& p-value  \\
		\hline
		
		water & 680&0.97 & 29&0.39 & 990&0.96 & 140&0.95& 14&0.19 & 240&0.80 \\
		
		1-butanol & 520 & 0.45& 59&0.27 & 2000&0.68 & 430&0.96 & 27&0.98 & 260&0.28 \\
		
		toluene & 600& 0.95 & 48 & 0.77 & 1200&0.87 & 190&0.89 & 46&0.38 & 300&0.46 \\
		
		cyclohexanone & 890 &0.94& 31&0.91 & 1250&0.79 & 395&0.37 & 110&0.61 & 437&0.96 \\
		
		cyclohexane & 540 &0.84& 55&0.37 & 1840 &0.84& 270&0.96 & 38&0.29& 320&0.51 \\
		
		acetonitrile & 1600 &0.97 & 60&0.91& 1570&0.65 & 267&0.80& 16&0.09& 330&0.89 \\
		
		trichloromethane & 2000 &0.59& 41&0.90 & 1400&0.63 &400&0.58 &11&0.48 & 420&0.96 \\
		\hline
		
		\textbf{state in the \{100\}} & \multicolumn{2}{c|}{\textbf{c1}} & \multicolumn{2}{c|}{\textbf{c2}} &\multicolumn{2}{c|}{\textbf{c3}} & \multicolumn{2}{c|}{\textbf{c4}} & \multicolumn{2}{c|}{\textbf{c5}} & \multicolumn{2}{c|}{\textbf{c6}}  \\
		\textbf{polar surface} & [ps] & p-value & [ps] & p-value &[ps] & p-value &[ps] & p-value &[ps] &p-value & [ps]& p-value  \\
		\hline
		
		water & 1760&0.91 & 91&0.97 & 218&0.20 & 90&0.72& 237&0.80 & 82&0.14 \\
		
		1-butanol & 1300 & 0.38& 131&0.75 & 1060&0.79 & 330&0.96 & 534&0.17 & 536&0.32 \\
		
		cyclohexanone & 1410 &0.37& 91&0.97 & 770&0.63 & 26&0.45 & 525&0.35 & 80&0.77 \\
		\hline
		\textbf{state in the crystal bulk} & \multicolumn{2}{c|}{\textbf{c1}} &&&&&&&&&& \\
		& [ps] & p-value &&&&&&&&&&\\
		\hline
		
		& 92 000&0.98  &&&&&&&&&&\\
		\hline
	\end{tabular}
	\label{surface_lifetimes}
\end{table*}

\subsection{Residence times of ibuprofen conformers}
The average residence time of each ibuprofen conformer in all relevant environments have been summarised in Tab. ~\ref{solution_lifetimes}, Tab.~\ref{adsorbed_lifetimes} and Tab. ~\ref{surface_lifetimes}. The latter includes the p-values for the KS tests used to validate the kinetic information recovered from WTmetaD. The lifetime of the conformers in each environment was used to set up the Markov State kinetic Models for the corresponding states.

\begin{figure}[h!]
	\includegraphics[width=1.0\linewidth]{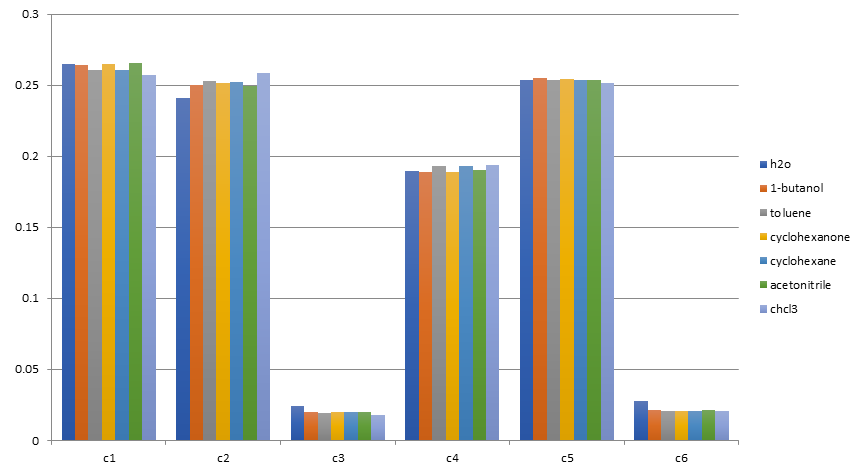}
	\centering
	\caption{Equilibrium distribution of ibuprofen conformers in bulk solution obtained with COSMO. }
	\label{COSMO}
\end{figure}

\subsection{Equilibrium Probability}
\begin{figure*}[h!]
	\includegraphics[width=0.5\linewidth]{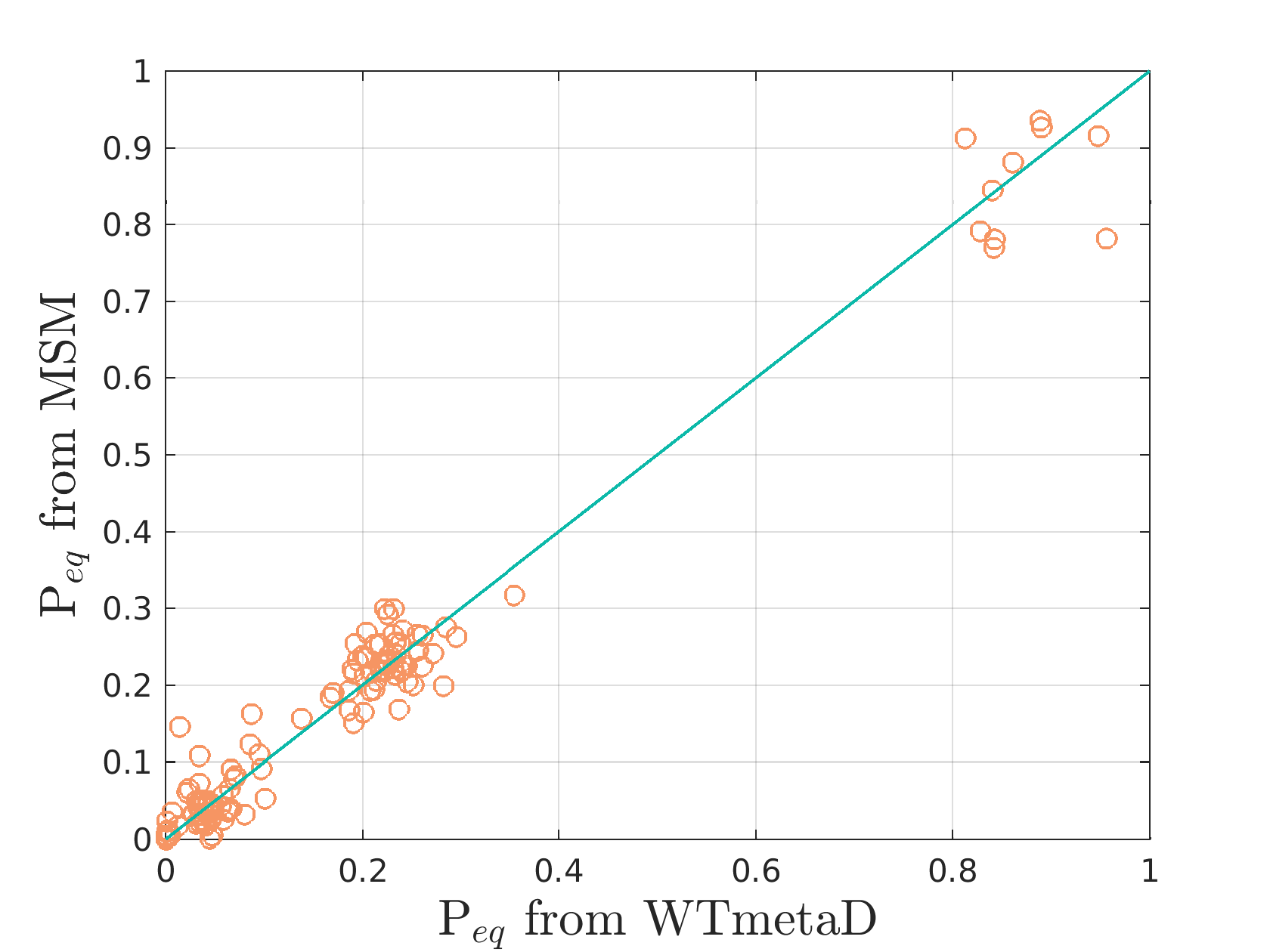}
	\centering
	\caption{Plot of equilibrium probability obtained from WTmetaD and MSM for ibuprofen conformers (orange) and the $x=y$ function (green).  }
	\label{PvsP}
\end{figure*}

The equilibrium probability recovered from the WTmetaD simulations in solution was verified against COSMO calculations as shown in ~\figurename{~\ref{COSMO}}. The Markov State Models, set up of each state, have been verified by comparing the equilibrium distribution obtained with the MSM with the corresponding WTmetaD one. The agreement between the two methos is shown in ~\figurename{~\ref{PvsP}}.

\begin{figure*}[h!]
	\includegraphics[width=0.9\linewidth]{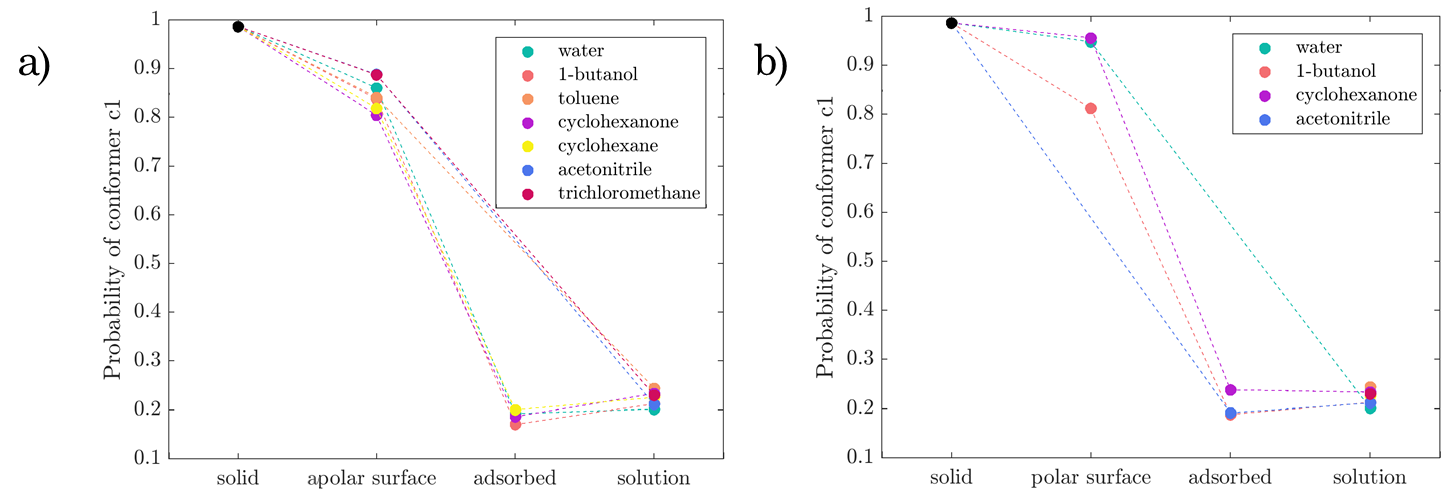}
	\centering
	\caption{Plot of \% of conformer c1,found in all different molecule environments and solvent, where (a) concerns the apolar \{100\} surface and (b) - the polar. }
	\label{fig:percent_c1}
\end{figure*}

\begin{figure*}[!]
	\makebox[\textwidth][c]{\includegraphics[width=1.2\linewidth]{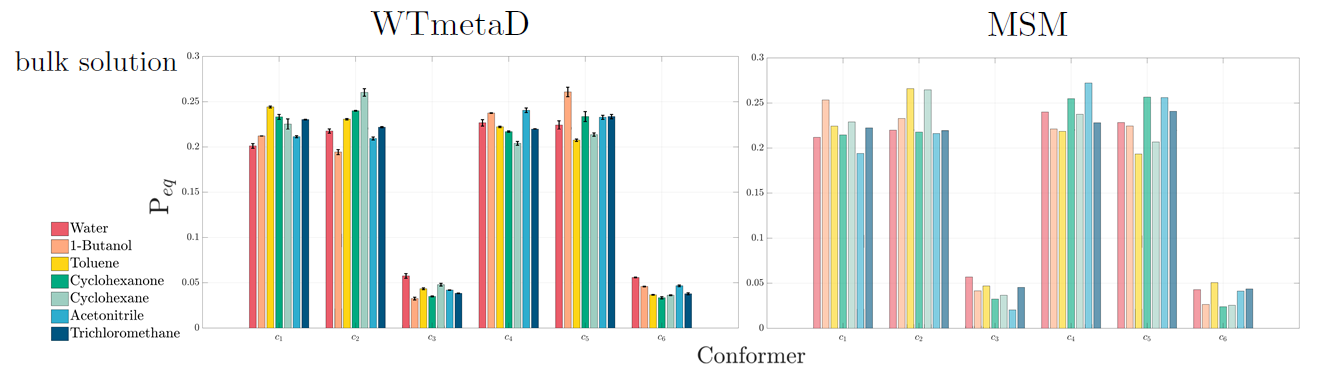}}
	\centering
	\caption{Probability of ibuprofen conformers in solution}
	\label{Solution}
	
\end{figure*}

\begin{figure*}[!]
	\makebox[\textwidth][c]{\includegraphics[width=1.\linewidth]{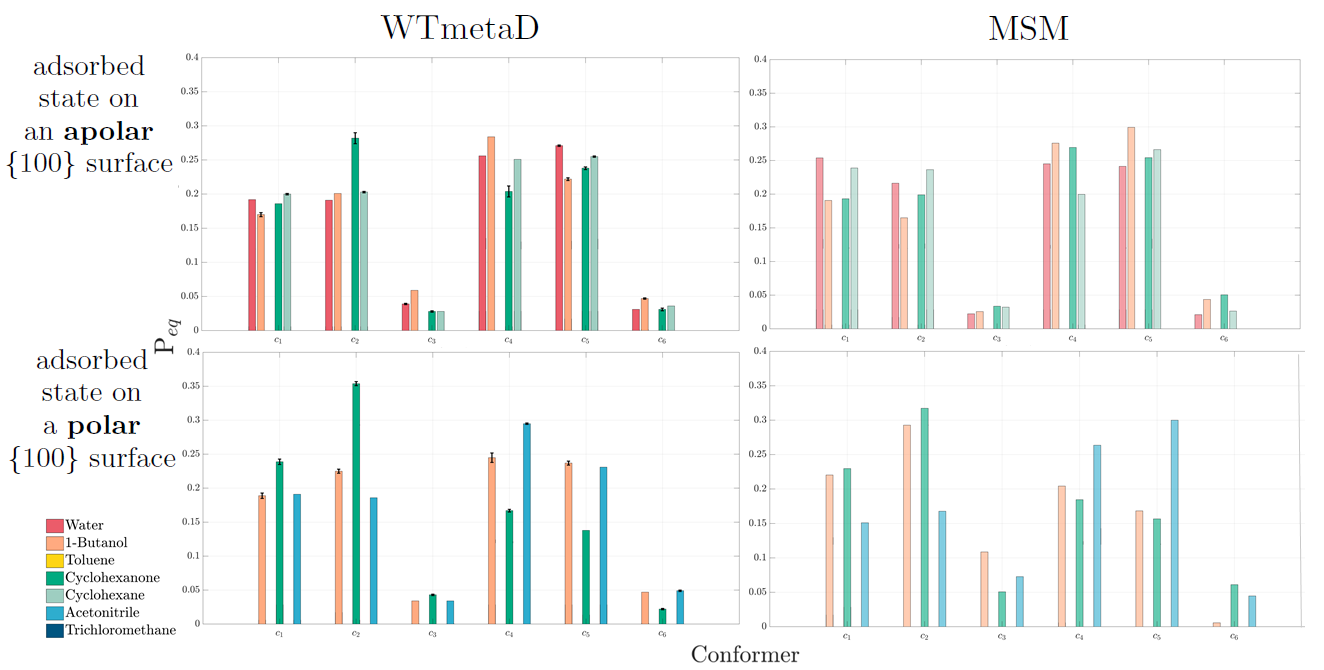}}
	\centering
	\caption{Probability of ibuprofen conformers in adsorbed states}
	\label{Adsorbed}
	
\end{figure*}

\begin{figure*}[!]
	\makebox[\textwidth][c]{\includegraphics[width=1.\linewidth]{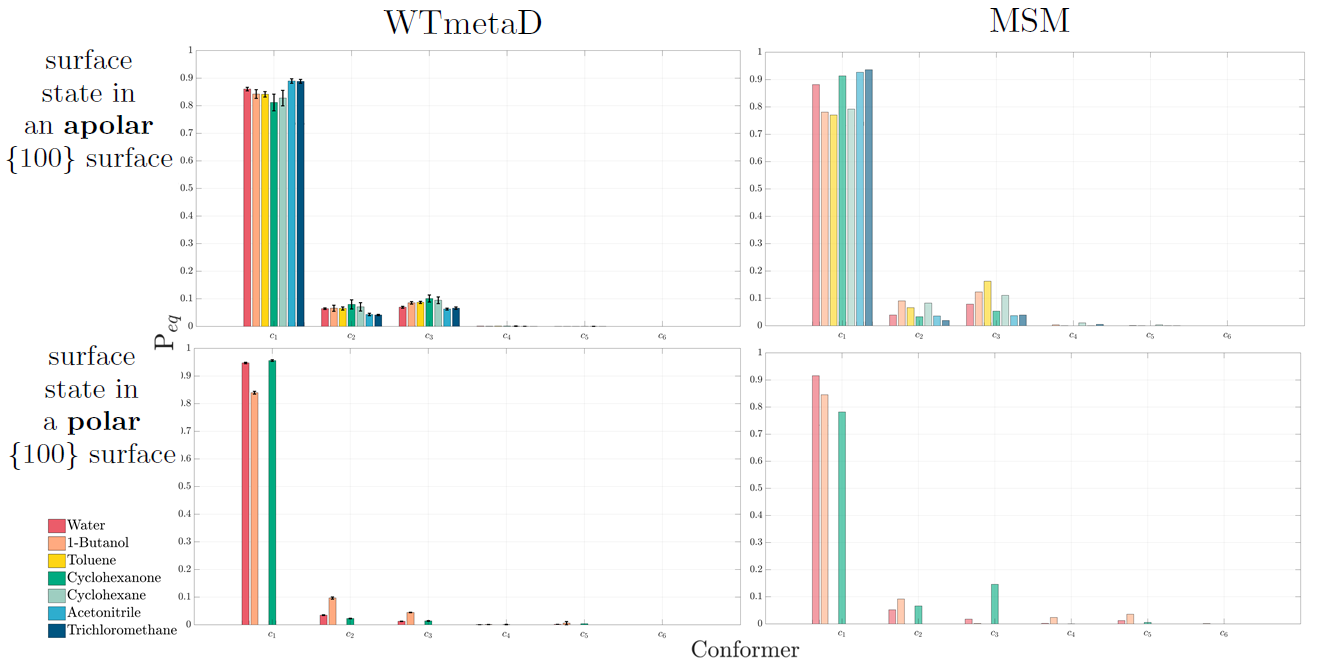}}
	\centering
	\caption{Probability of ibuprofen conformers in the surface states}
	\label{Surface}
	
\end{figure*}

The percent c1 conformer found in each case discussed can be found in Fig ~\ref{fig:percent_c1}. The equilibrium probability for all conformers in all discussed cases, recovered from WTmetaD and MSM, can be found in ~\figurename{~\ref{Solution}},~\figurename{~\ref{Adsorbed}} and 
~\figurename{~\ref{Surface}}. The probability of the conformers recovered from WTmetaD, and the associated error, has been calculated using the last 30 \% of each simulation for convergence reasons.

\begin{figure*}[!]
	\makebox[\textwidth][c]{\includegraphics[width=1.\linewidth]{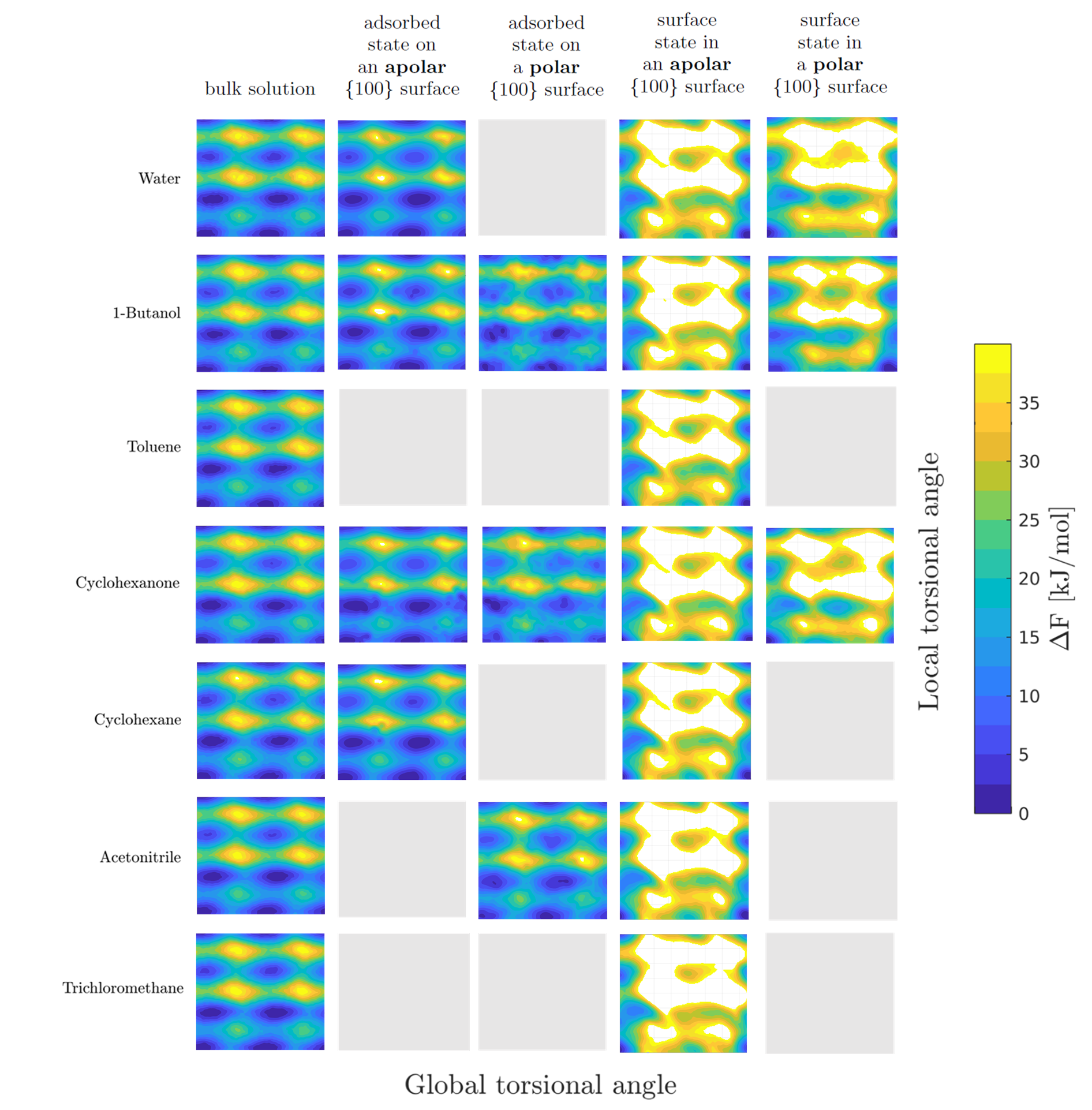}}	\centering
	\caption{Free Energy Surface Summary}
	\label{FES}
\end{figure*}

\begin{figure*}[!]
	\makebox[\textwidth][c]{\includegraphics[width=1.\linewidth]{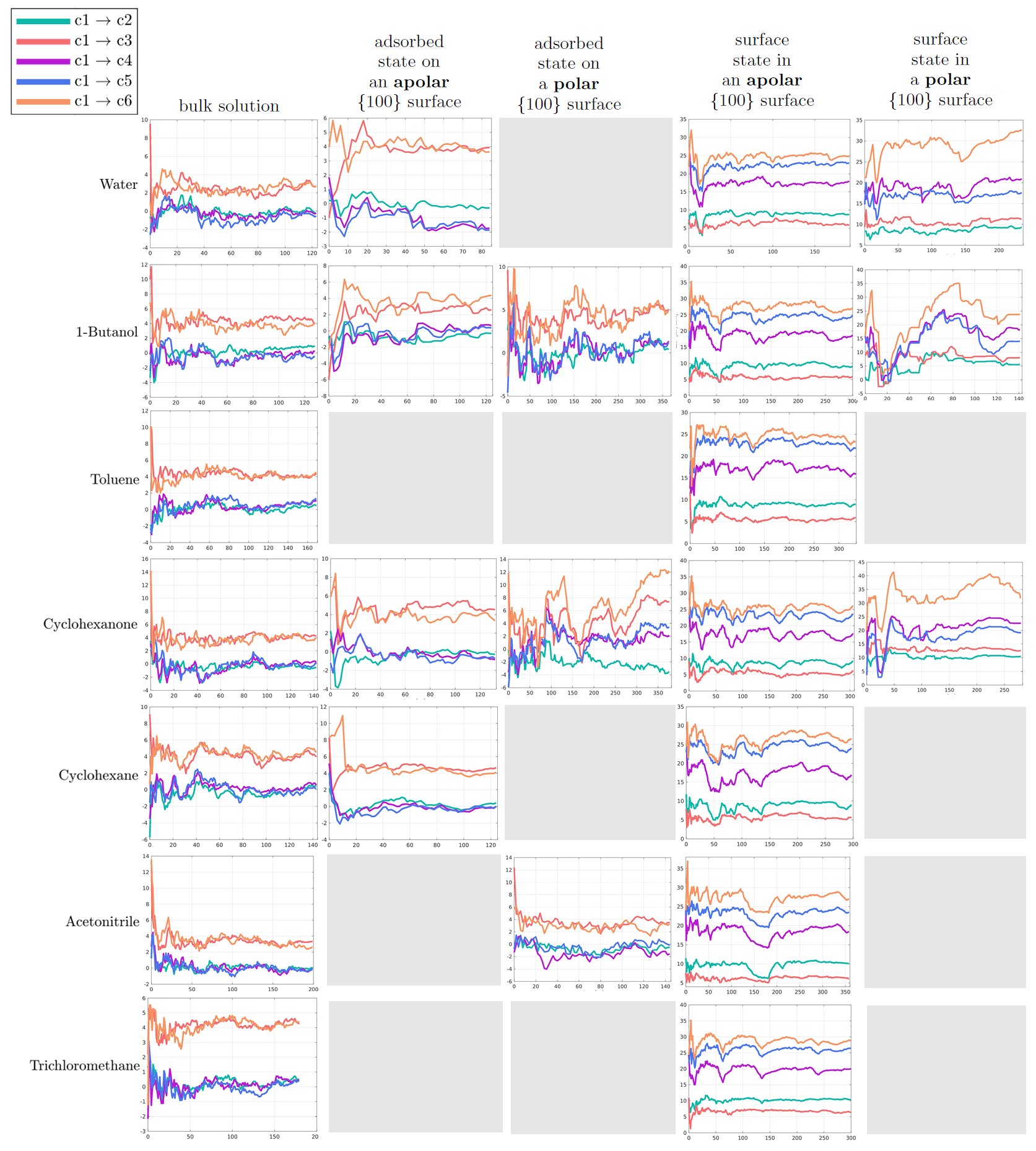}}	\centering
	\caption{Convergence Summary}
	\label{Convergence}
	
\end{figure*}

\figurename{~\ref{FES}} and \figurename{~\ref{Convergence}} provide the free energy surface and convergence plots for all cases.


\end{document}